\documentclass[10pt,a4paper,twocolumn]{article}
\usepackage[utf8]{inputenc}
\usepackage[T1]{fontenc}
\usepackage[top=0.75in,bottom=0.75in,left=0.6in,right=0.6in,headheight=20pt]{geometry}
\usepackage{graphicx}
\usepackage{booktabs}
\usepackage{array}
\usepackage{xcolor}
\usepackage{hyperref}
\usepackage{amsmath,amssymb}
\usepackage{enumitem}
\usepackage{float}
\usepackage{titlesec}
\usepackage{fancyhdr}
\usepackage{caption}
\usepackage{xurl}
\usepackage{colortbl}
\usepackage{tabularx}
\usepackage[numbers,sort&compress]{natbib}

\definecolor{servablue}{RGB}{30,60,114}
\definecolor{servalight}{RGB}{235,242,250}
\definecolor{servagold}{RGB}{184,134,11}
\definecolor{servagray}{RGB}{100,100,100}
\definecolor{lightgray}{RGB}{248,248,248}
\definecolor{accentgreen}{RGB}{46,125,50}

\hypersetup{
    colorlinks=true,
    linkcolor=servablue,
    citecolor=servablue!70,
    urlcolor=servablue!80,
    pdftitle={The .serva Standard: One Primitive for All AI},
    pdfauthor={Servamind Inc.}
}

\titleformat{\section}{\large\bfseries\color{servablue}}{\thesection.}{0.5em}{}[\vspace{0.1em}\textcolor{servablue!30}{\rule{\columnwidth}{0.5pt}}]
\titleformat{\subsection}{\normalsize\bfseries\color{servablue!85}}{\thesubsection}{0.5em}{}
\titleformat{\subsubsection}{\small\bfseries\color{servablue!75}}{\thesubsubsection}{0.4em}{}

\titlespacing*{\section}{0pt}{1.8ex plus 0.5ex minus 0.2ex}{1.0ex plus 0.2ex}
\titlespacing*{\subsection}{0pt}{1.4ex plus 0.3ex minus 0.1ex}{0.6ex plus 0.1ex}

\setlist{nosep,leftmargin=1.2em,itemsep=0.15em}

\renewcommand{\headrulewidth}{0.4pt}
\renewcommand{\headrule}{\hbox to\headwidth{\color{servablue!40}\leaders\hrule height \headrulewidth\hfill}}

\newcommand{\serva}{\texttt{.serva}}
\newcommand{\keyinsight}[1]{%
\vspace{0.3em}
\noindent\fcolorbox{servablue!50}{servalight}{\parbox{\dimexpr\columnwidth-2\fboxsep-2\fboxrule}{\small\textit{#1}}}
\vspace{0.3em}
}
\newcommand{\metric}[1]{\textbf{\textcolor{servablue}{#1}}}
\newcommand{\highlight}[1]{\colorbox{servagold!15}{\textbf{#1}}}

\newcommand{\sectionrule}{\vspace{0.2em}\noindent\textcolor{servablue!30}{\rule{\columnwidth}{0.3pt}}\vspace{0.2em}}

\begin{document}

\twocolumn[
\begin{@twocolumnfalse}
\vspace*{-0.3in}

\noindent\textcolor{servablue}{\rule{\textwidth}{1.5pt}}
\vspace{0.15in}

\begin{center}
{\LARGE\bfseries\textcolor{servablue}{The .serva Standard: One Primitive for All AI}\par}
\vspace{0.12in}
{\large\textcolor{servablue!70}{Cost Reduced, Barriers Removed}\par}
\vspace{0.18in}
{\normalsize\textbf{Servamind Inc.} --- \textit{Part One}\par}
\vspace{0.08in}
{\small Rachel St.~Clair, John Austin Cook, Peter Sutor Jr., Victor Cavero, Garrett Mindt\par}
\vspace{0.06in}
{\small\href{https://servamind.com}{servamind.com} ~|~ \href{mailto:info@servamind.com}{info@servamind.com} ~|~ December 2025\par}
\end{center}

\vspace{0.15in}
\noindent\textcolor{servablue}{\rule{\textwidth}{0.5pt}}
\vspace{0.15in}

\noindent\textbf{\textcolor{servablue}{Abstract}} --- Artificial Intelligence (AI) infrastructure faces \textbf{two compounding crises}. \textit{Compute payload} --- the unsustainable energy and capital costs of training and inference threatens to outpace grid capacity and concentrate capability among a handful of organizations. \textit{Data chaos } --- the 80\% of project effort consumed by preparation, conversion, and preprocessing --- strangles development velocity and locks individual datasets to single model architectures. Current approaches treat these as separate problems, managing each with incremental advancements in optimization, increasing complexity in the overall AI tooling ecosystem. The approach presented here views data and computation as two expressions of single architecture, where a unified primitive is missing. This paper presents \textbf{ServaStack}: a universal data format (\serva) paired with a universal AI compute engine (\textbf{Chimera}). The \serva{} format achieves lossless compression by encoding information using laser holography principles, while Chimera converts compute operations into a representational space where computation occurs directly on \serva{} files without decompression. For AI, the result is automatic data preprocessing by converting into \serva. The Chimera engine enables any existing model to operate on \serva{} data without retraining, preserving infrastructure investments while revamping their efficiency. Internal benchmarks demonstrate \metric{30--374$\times$ energy efficiency} improvements (96--99\% energy reduction), \metric{4--34$\times$ lossless compression}, and \metric{68$\times$ compute payload reduction} without loss of accuracy when compared to RNN, CNN, and MLP models trained on original FashionMNIST and MNIST dataset files and compression compared to popular SOTA compression algorithms on Canterbury Corpus. At hyperscale one billion daily iterations, these gains translate cost savings of \$4.85M per petabyte per training cycle. The impact of this technology proposes more significance than the efficiency; that when any data flows to any model on any hardware, the entire AI development paradigm shifts. The bottleneck moves from infrastructure to imagination.

\vspace{0.12in}
\noindent\textcolor{servablue}{\rule{\textwidth}{0.5pt}}
\vspace{0.2in}
\end{@twocolumnfalse}
]

\section{Introduction}

\textit{Any data to any model on any hardware.} This is what Servamind has built.

The history of computers is littered with examples of advances which we scarcely remember now, but fundamentally changed the way we use, develop, and deploy new computer or AI technology. Almost imperceptible today they laid the groundwork for the ecosystem we now work, develop, deploy, and occupy. Just as \textbf{Grace Hopper} unlocked code to machine program lock-in, we are in the era of programming with \textbf{data to model lock-in}. AI lacks a primitive that removes the need for data preparation to be hand-crafted with the downstream model in mind---similar to the pre-Grace Hopper programmers who had to detail compute instructions directly to the specific target machine. With these kinds of advancements we usher in new paradigms of development, opening up new possibilities for both developers and consumers. There are two problems holding back bringing the next era of AI development to fruition and that is the \textbf{compute payload} required to train foundational and frontier models and the \textbf{data chaos} involved in curating data to train those same models. Servamind has developed a solution to these problems with our \textbf{Serva Encoder} and our AI wrapper, \textbf{Chimera}. With these two solutions anyone can take any data, in any format, and train any model on any hardware. With Servamind we open up a new world of possibilities for the next generation of AI development.     

AI infrastructure spending reached \textbf{\$135 billion in 2024} and is projected to surpass \$200 billion by 2028 and the figure is accelerating~\cite{ref16}. Yet AI remains trapped: training accessible only to hyperscalers, impractical at the edge, unsustainable at scale. The cardinal constraint is feasibility, which relies on infrastructure, encouraging new forms of machine intelligence to emerge.

Servamind's purpose is to allow AI to flourish in a meaningful way to all of society. To achieve this we have built \textbf{Servastack}, a combination of our unique encoder (Serva Encoder) and an AI wrapper (Chimera) to solve what we think are two difficult problems in the AI development pipeline: compute payload and data chaos. Servastack is a \textit{universal data format paired with a universal AI compute engine}, allowing data and computation to flow across any ecosystem into a unified application. This property of universality is paired with an exploit on the current compute paradigm for maximum efficiency. This brings down the cost of AI creation and use by orders of magnitude. Rather than copying and optimizing current approaches, the focus of AI advancement can now become new modes of learning, new user interfaces, and new insights into thinking machines.

\subsection{The Two Primary Problems}

AI cost is dominated by two challenges: \textbf{data chaos} and \textbf{compute payload}. In plain terms, \textit{AI is expensive and data is difficult}. Together, these account for the vast majority of project resources. Data preparation alone consumes roughly \textbf{80\% of total effort and budget}, while compute infrastructure represents 47--67\% of total AI development costs for organizations building from scratch~\cite{ref8,ref18}.

\sectionrule

\noindent\textbf{\textcolor{servablue}{Compute Payload}} is the visible crisis. The International Energy Agency projects that global datacenter electricity consumption will more than double by 2030, reaching approximately \metric{945 TWh annually}---equivalent to Japan's entire electricity consumption~\cite{ref1}. AI-accelerated servers are growing at 30\% annually, four times faster than total electricity supply growth, and 20\% of planned datacenter projects already face delays due to grid bottlenecks. Grid operators are already hitting capacity limits: in Virginia, Dominion Energy faces a \textbf{seven-year backlog} for new datacenter connections; Ireland imposed a moratorium on Dublin datacenter grid connections from 2021 to 2025 due to electricity system strain~\cite{ref19,ref20}. The U.S. Department of Energy warns that without efficiency breakthroughs, AI's power demands could require dozens of new power plants within the decade~\cite{ref2}.

The scale of investment often reflects the scale of demand:

\begin{table}[H]
\centering
\caption{AI Infrastructure Investment Scale (2024--2025)}
\label{tab:investment-scale}
\small
\begin{tabular}{@{}lr@{}}
\toprule
\textbf{Indicator} & \textbf{Value} \\
\midrule
NVIDIA datacenter revenue & \$15B $\rightarrow$ \$47B~\cite{ref3} \\
Microsoft AI datacenter commitment & \$80B (2025)~\cite{ref4} \\
Frontier compute growth & 4--5$\times$/year~\cite{ref5} \\
GPT-4 training cost & $\sim$\$78M~\cite{ref6} \\
Gemini Ultra training cost & $\sim$\$191M~\cite{ref6} \\
GPT-4 CO$_2$ emissions & 10--15K metric tons~\cite{ref7} \\
\bottomrule
\end{tabular}
\end{table}

ARK Invest offers a counternarrative: AI training costs are declining approximately 75\% annually through Wright's Law dynamics, hardware improvements reducing compute unit costs by 53\% per year, compounded by algorithmic efficiencies contributing another 47\%~\cite{ref22}. If costs decline accordingly, the argument goes, AI scales sustainably.

This analysis, however, \textbf{conflates efficiency with capability}. Wright's Law measures the cost to reproduce yesterday's performance, not to achieve tomorrow's, not to increase AI efficacy (i.e.\ intelligence). Moreover, ARK's cost curves track per-unit compute while excluding the infrastructure buildout where costs are rising and resources are constrained: datacenter real estate prices increased 19\% in 2024, supply chain shortages for generators, chillers, and transformers are inflating construction costs, and grid connection backlogs stretch to seven years in key markets---none of which follow Wright's Law dynamics~\cite{ref29,ref30,ref31}. The view that AI is scaling is myopic compared to the total supply chain growth needs of current AI's trajectory.

DeepMind's \textbf{Chinchilla scaling laws} reveal the deeper constraint in large language model (LLM) development. The relationship between compute and capability follows a power law~\cite{ref23}. Compute-optimal training requires scaling parameters and data together, with FLOPs scaling quadratically with model size ($\approx 6ND$ for dense transformers) while capabilities improve along a much shallower curve. The implication is sobering: \textit{frontier capability does not get cheaper}. Each incremental improvement in model performance demands disproportionately more compute. The goalpost moves faster than efficiency gains can follow.

The consequences are threefold: (1) \textbf{Climate impact} is mounting. (2) \textbf{Capability is concentrating}---only a handful of organizations can afford frontier model training. (3) \textbf{Barriers to entry} are rising; startups, researchers, and developing nations find themselves increasingly locked out of meaningful participation in AI advancement. The Chinchilla constraint compounds all three: every capability improvement demands proportionally more data, more compute, and more energy. Yet, Chinchilla applies only to LLMs. The next frontier in generative AI---simulation models, multi-modal systems, and multimedia reasoning generation---exhibits even steeper scaling requirements. Section~\ref{sec:cost} examines how Servamind's efficiency gains alter this calculus across modalities.

\textbf{AI in its current form is not scalable to the degree that its collateral costs should be overlooked.}

\sectionrule

\noindent\textbf{\textcolor{servablue}{Data Chaos}} is the less visible crisis even though AI developers live it daily.

The practical experience goes as follows: spend months determining how to prepare data and then quickly copy/paste a model architecture from a journal repository. Refit this model in a week. Then run inference for another week, before finally obtaining some results. Then after all of that, refactor everything for a new model and repeat the entire cycle. \textit{The actual AI is the easy part. The real job is all the data work.}

Industry analyses consistently estimate that \textbf{80\% or more} of any AI project's effort goes to data preparation, cleaning, and orchestration~\cite{ref8}. This reality is reflected in market behavior: models are frequently open-sourced while training data remains sacred, proprietary, high-priced IP. The value resides in the data, and the cost resides in preparing it.

There is also a \textbf{one-to-one lock-in between dataset and model}. Data cannot simply be fetched from storage and passed to any AI. It must be specifically pre-processed for the downstream AI model architecture. Every time a new model is adopted, the data must be re-processed from scratch to produce model identifiable features. Aside from being inefficient, it is genuinely frustrating for practitioners who understand that the underlying information remains the same regardless of which model will consume it. Feature engineering is a human operation; eliminating it would allow AI models to do the bulk of the work.

As data capture expands, the problem compounds. IDC projects global data creation will exceed \textbf{180 zettabytes by 2025}, up from 64 zettabytes in 2020~\cite{ref9}. Much of this growth comes from specialized domains---medical imaging, satellite telemetry, genomics, industrial sensors---producing formats poorly matched to mainstream architectures. Regulated industries like healthcare and finance face 20--35\% higher AI implementation costs due to compliance requirements, specialized data handling, and domain adaptation needs~\cite{ref21}. The prevailing data constraints are format issues, quality issues, scale issues, and security issues. 

\textbf{The data problem is accelerating, not slowing.}

\subsection{Current Approaches}

\textbf{Hardware approaches} are largely abetted by miniaturization, making transistors smaller so more of them can fit on chip. Moore's Law, the observed rate of miniaturization, however, is lessening. Perhaps 10--15 years of conventional scaling remain before atomic limits impose quantum effects~\cite{ref10}. Practical quantum computing at consumer scale is not expected until 2040 or beyond~\cite{ref11}. The hardware path alone will not solve the timeline we face. 

\textbf{Software approaches} seek efficiency through representation: 64-bit to 8-bit quantization, pruning, distillation, sparsity. Even aggressive techniques like Microsoft's BitNet, which reduces weights to ternary values \{-1, 0, +1\}, achieve 2--6$\times$ speedups and 55--82\% energy reduction~\cite{ref24}. In an attempt to overcome data chaos, typical approaches are to orchestrate systems and layers to navigate the chaos. Some current approaches attempt data orchestration: Pandas DataFrames, PyTorch tensors, Hugging Face SafeTensors. ML-Ops platforms like MLflow, Weights \& Biases, and Kubeflow coordinate infrastructure and track experiments. 

\textbf{Data format approaches}: The AI ecosystem has also produced numerous data formats, each solving a narrow problem. Apache Arrow and Parquet optimize columnar analytics but assume tabular structure. TFRecord and SafeTensors serialize tensors for specific frameworks. ONNX provides model interchange but not data interchange and remains a conversion layer, not a native format. 

Current approaches represent steps in the right direction towards unifying the AI ecosystem's tooling. Yet \textbf{none are universal for all data to any model}. Further, they do not address the root cause of binding feature information to compute payload. While these approaches deliver valuable gains, none deliver the order-of-magnitude improvements required to break scaling constraints, or ease the data-model lock-in. Each standard addresses one link in the pipeline while the fundamental fragmentation remains. 

The AI field moves too rapidly to design for specific configurations. It is futile for any team to keep pace with every variation. Too many data formats exist. Too many model architectures compete. Too many hardware targets fragment the landscape. Too many programming languages divide practitioners. Tooling is scattered, redundant, and confusing. 

Rather than spend so much time navigating the chaos we set out to solve the problem by simplifying all data preprocessing using a universal encoding producing a standard file format: a \textbf{\serva{} file}.

\sectionrule

\noindent\textbf{\textcolor{servablue}{How .serva Differs from Existing Standards.}}
The \serva{} format differs \textit{in kind, not degree}. It is not a serialization format for a specific data type. It is not a conversion layer between frameworks. It is a \textbf{universal encoding} that transforms any input (images, text, audio, sensor streams, structured records, etc.) into a single representational space where all information is preserved and direct information extraction (i.e.\ computation) can occur. The question shifts from ``how do I convert my data for this model'' to ``\textbf{encode once, compute anywhere}.''

\begin{table*}[t]
\centering
\caption{Traditional Data Pipeline vs.\ .serva Encoding. Standard AI data preparation requires six distinct steps, each demanding specialized tooling and domain expertise. The .serva format collapses this pipeline into a single encoding step, with Chimera handling model integration. Steps that traditionally consume 80\% of project effort become automatic properties of the representation.}
\label{tab:pipeline-comparison}
\begin{tabular}{@{}p{2.2cm}p{6.5cm}p{6.5cm}@{}}
\toprule
\textbf{Pipeline Step} & \textbf{Traditional Approach} & \textbf{.serva Approach} \\
\midrule
\textbf{Validation} & Manual error detection and correction for impossible values, type mismatches, and date errors. Requires domain-specific rules and quality assurance passes. & High-dimensional encoding is robust to sparse errors; anomalies do not significantly impact downstream processing. \\
\addlinespace
\textbf{Format Conversion} & Convert between CSV, JSON, PDF, JPEG, TIFF, HEIC, etc. Write custom parsers per format. Maintain conversion code as formats evolve. & Single encoder accepts any input format; outputs universal .serva representation regardless of source format. \\
\addlinespace
\textbf{Cleaning} & Remove duplicates, standardize dates, strip whitespace, normalize capitalization, handle missing values. Often 30--50\% of preparation time. & Encoding normalizes representations automatically; surface-level variations map to similar vector representations. \\
\addlinespace
\textbf{Feature Engineering} & Domain experts manually extract and select relevant features. Must be repeated for each new model architecture. & Lossless encoding preserves all features; downstream model extracts what it needs via Chimera wrapper. \\
\addlinespace
\textbf{Data Augmentation} & Apply random transforms (rotation, crop, flip, color jitter) to increase diversity. Requires framework-specific implementations. & Augmentation can be performed directly in encoded space through vector operations. \\
\addlinespace
\textbf{Data Loading} & Batch management, shuffling, GPU memory optimization. Framework-specific loaders (PyTorch DataLoader, tf.data, etc.). & Chimera handles batching and device placement; compressed format reduces memory and transfer overhead. \\
\bottomrule
\end{tabular}
\end{table*}

\subsection{Servamind Approach}

The root inefficiency cause is that data and computation have never been addressed together in a way that leverages the existing ecosystem. Our answer to these two compounding problems is our \textbf{Servastack}, a universal data format (\serva) to eliminate data chaos and a universal AI compute engine (Chimera) that tackles the compute payload problem. These two solutions are independently needed in order to penetrate the existing ecosystem at various levels of the pipeline, where the two problems remain separate. When combined in a workflow, Servastack emerges as a unification paradigm that can begin to be leveraged for compounding efficiency gains. 

\textbf{Servamind attacks both problems simultaneously through a unified system:}
\begin{itemize}
\item A \textbf{universal data format} (\serva) that eliminates data chaos, created by Serva Encoder  
\item A \textbf{universal compute engine} (Chimera) that solves compute payload  
\item \textbf{Together (Servastack)}: 30--374$\times$ energy efficiency (96--99\% cost reduction), $\sim$4$\times$ lossless storage compression, and 34$\times$ compute payload reduction---without diminished accuracy   
\item \textbf{Data agnostic. Model agnostic. Hardware agnostic.}
\end{itemize}

\keyinsight{The insight most observers miss: they hear ``compression'' and think ``efficiency.'' Efficiency is a consequence. \textbf{Universality is the breakthrough.}}

When any data can flow to any model, the entire AI development paradigm shifts so that any model can ingest any data. \textbf{Bidirectional compatibility} emerges. The data layer becomes decoupled from the model layer entirely. True multimodality becomes tractable---vision-language-action models in robotics have struggled not for algorithmic reasons, but because fitting heterogeneous data into one model presents an engineering nightmare. The data-model lock dissolves across all verticals.

The market reality shaped the product. In the current AI ecosystem only hyperscalers can afford frontier training. The efficiency gained through Servastack would democratize access. But adoption faces a barrier, since organizations resist retraining models in which they have already heavily invested. Infrastructure overhaul appears more expensive than ongoing inefficiency.

This constraint forced two requirements:
\begin{enumerate}
\item More efficient computation regardless of data type or source  
\item Compatibility with any stage of AI-training, pre-training, fine-tuning, inference 
\end{enumerate}

What is most important to note: \textbf{the Servamind solution does not contradict or compete with other approaches}. It is universal. It is additive. 

\keyinsight{Servastack partners with and amplifies the efforts of all those pursuing ease and efficiency in AI.}

\section{The Origin}

Servamind began with a question about why AI fails to scale like biological intelligence.

The culprit in neural networks is \textbf{catastrophic forgetting}, known as the fundamental limitation of backpropagation-based learning. Alternative paradigms hit their own walls: reinforcement learning's sample inefficiency, knowledge graphs' combinatorial explosion, symbolic AI's brittleness~\cite{ref25,ref26,ref27,ref28}. No existing approach scales cleanly. The workarounds remain fragile, and the industry locks in technical debt with every deployment. When a neural network learns a new task, it updates its weights to optimize for that task at the expense of prior tasks. The model drifts away from previous knowledge, constantly forgetting what it once knew. This is not a bug in implementation; it is a structural consequence of how greedy-based learning operates~\cite{ref12}.

Brains learn a bit differently. Biological neural systems do not usually catastrophically forget. A human who learns Spanish does not entirely lose their English for doing so, as would, for example, Siri's AI. The brain has evolved to solve this problem. This recognition initiated a search for learning mechanisms closer to biological reality.

\subsection{Inspiration from Biological Systems}

That search led to the work of \textbf{L.\ Andrew Coward}, namely \textit{Recommendation Architecture}, a theoretical framework for understanding higher cognition in terms of anatomy and physiology~\cite{ref13}. Years of study revealed several foundational insights that would reshape our approach to the problem.

\textbf{First}: Information in biological systems is computed in \textit{three-dimensional physical space}. The spatial arrangement of neurons matters. This makes von Neumann architectures, where memory and processing are fundamentally separated, a poor substrate for brain-like computation. Every mainstream computer inherits this bottleneck.

\textbf{Second}: The units of information in biological systems are \textit{maximally combinatorial}. They are designed to combine and build up into any higher-order representation for unknown future tasks. The brain does not optimize its representations for the current task; it maintains flexibility for tasks it has never encountered. Representations are distilled into suggestions that the system learns from, not hard commitments that foreclose future possibilities.

\textbf{Third}: The filtering of noise to signal does not happen inside the brain the way it happens inside AI models. In traditional AI, feature engineering and model layers perform noise-to-signal transformation. In biological systems, however, that filtering has already occurred \textit{upstream, at the sensor}. The retina, the cochlea, and the mechanoreceptors are not passive recorders. They are intelligent filters shaped by hundreds of millions of years of evolution. By the time information reaches the brain, coherence has already been imposed by the camera, the lidar sensor, the capture device.

Representations should \textbf{preserve possibility, not collapse it prematurely}---like the relationship between DNA and protein, where the same genetic sequence can participate in producing vastly different outcomes depending on context. This implies that internal representations should be more ambiguous.

How can we know what information will matter when the downstream task is unknown, as is often the case in general intelligence and in practical AI development? The approach Servamind takes is to \textbf{preserve everything} and assume all apparent noise may be a latent signal. We keep everything, while increasing efficiency as a byproduct. In later sections, we explain how this paradox is resolved. 

\subsection{Compression and Intelligence: The Hutter Framework}

\textbf{Marcus Hutter's} work from Google DeepMind establishes a profound equivalence: \textit{compression and prediction are fundamentally the same operation}~\cite{ref14}. To compress data optimally is to model it optimally. Optimal compression implies optimal inference. ``If you can compress, you can learn'' is a mathematical identity rooted in Kolmogorov complexity and algorithmic information theory~\cite{kolmogorov1965}.

The implication for AI systems is significant: \textbf{superior compression yields superior efficiency}. A system that achieves better compression has, by definition, extracted more structure from data using fewer resources. Most approaches follow the trend of incremental optimization. Our approach adds to incremental optimization by providing a primer designed from exploitation of mathematics about the nature of learning itself.

In practical terms, successful compression separates signal from noise in a useful way. In AI, we call this \textit{feature engineering} and it is often performed by expert-crafted reduction operations to reduce unnecessary information in the data. The feature engineered data is then ingested by the AI model. The outputs of intermediate model network layers, before final classification or generation, are \textbf{feature vectors}: compressed representations that capture learned structure by losing the unlearned structure. The unlearned structure, or appropriate information to lose throughout the feature vector creation process is directly determined by its unnecessity to produce correct results for the desired learning task at hand. \textit{Compression and learning are the same operation viewed from different angles.}

\subsection{Information Theory: The Shannon Foundation}

\textbf{Claude Shannon's} foundational work on noisy channels established the theoretical limits of information transmission~\cite{ref15}. His central insight: information can be preserved through transformation if entropy is managed correctly. There exist transformations that reduce representation size while losing nothing---the domain of \textbf{lossless compression}.

Shannon's framework offers an additional insight relevant to our problem. Information was defined, in part, as non-randomness: structure, pattern, predictability. Determining whether apparent randomness is true stochasticity or merely a temporal state in an intractable deterministic system is not possible without complete observation. Information is thus defined in juxtaposition to noise. Whatever is not useful, we call noise. Usefulness, however, is task-dependent and often unknowable in advance.

The distinction between \textbf{lossy} and \textbf{lossless} compression matters critically for AI. Lossy compression discards information deemed unimportant by some prior criterion. In AI applications, however, we often cannot know in advance what information will prove important for downstream tasks. Lossless compression preserves all information, deferring the question of relevance to the learning system itself.

\subsection{The Paradox}

These frameworks created a \textbf{paradox} at their intersection.

\textbf{Hutter says}: compress to learn since optimal compression implies optimal prediction. AI systems should aggressively compress their representations to maximize learning efficiency, which by current methods involves drastic information reduction, further indebting the data-model lock.

\textbf{Shannon says}: preserve to remain general. Discarding information precludes possibilities. However, retaining all information would explode the compute resources required. The Shannon insight points in a possible direction for a solution: lossless compression. For tasks where downstream use is unknown, lossless preservation is required. 

These imperatives seem opposed. \textit{Compression discards; preservation retains.} How can both be satisfied simultaneously?

\section{Our Solution}

The resolution came from recognizing \textbf{where entropy actually exists} in the pipeline.

Shannon's framework assumes a noisy channel---a transmission medium that introduces randomness. The data AI systems operate on, however, is not raw entropy. It is \textit{captured data}: images from cameras, audio from microphones, telemetry from sensors, text from human authors.

These capture devices were designed by human intelligence. They impose coherence. They select what to record and how to record it. The camera does not capture pure photon chaos; it captures structured light filtered through a lens system engineered for human visual understanding. By the time data enters an AI pipeline, it has already been filtered by the causal structure of physical reality and the intentional design of the capture mechanism.

\textbf{The entropy is already reduced.} The signal has already been extracted, not by the AI, but by the physical and engineered systems upstream. Physical reality is causal---each state follows coherently from the last. Capture devices record this coherence. The data we compress is not noise. It is structured by physical causality and human intent. We are not fighting entropy---we are revealing the structure that was always present. 

Hutter established that compression enables learning. Shannon established how to compress without losing information. \textbf{Both can be satisfied simultaneously} when the data is already structured---and real-world data is structured by physical causality and human intent.

Servamind applies this synthesis: a lossless compressed format that AI models can compute on directly. Not lossy approximation, not dimensionality reduction---\textbf{lossless compression that retains signal} because, when downstream tasks are unknown, all of it may be signal.

\subsection{Universal Feature Vectors}

This raises a practical question: if we want feature vectors from data yet do not know what downstream model or task will consume them, how do we know what to encode?

The answer follows from the theory above: \textbf{encode everything}. Preserve all structure. Let downstream tasks extract what they need rather than guessing in advance what they will require. The difficulty of solving the problem then lies the challenge of \textit{how}---how to create lossless compression across arbitrary information, which is addressed partially in the next section on Architecture. 

Servamind encodings (\serva{} files) are representations that \textbf{preserve information}, because reduction may damage what matters. They trend towards being \textit{maximally combinatorial}, able to serve any downstream task because they have not been optimized for any particular one. 

This framing also addresses catastrophic forgetting at its source. Backpropagation-based learning overwrites weights optimized for previous tasks when learning new ones. The model constantly drifts from prior knowledge. Learning on universal feature vectors, however, provides protection at the data layer. Information is not discarded because all of it is treated as information. The representations themselves resist the forgetting problem by never having collapsed the possibility space in the first place, as long as the computation interacts in this encoded space.

\subsection{Why No One Attempted this Until Now}

In short, this challenge is a monumental opportunity, where many other lower hanging fruits are to be found. The paradox of keeping everything and increasing efficiency also sounds implausible on its face. Compression and computation appear to be opposing operations. Compression standards are focused on removing noise to filter the data for easier computation. Lossless compression lacked any robust implementation that could rival lossy methods. Further computing on any lossless compression is narrow and few implementations exist at all, let alone practically and market viable ones.

This apparent contradiction dissolves under a different computational paradigm. Compression and computation seem opposed only when data is lost or must decompress before processing. \textbf{When the encoding itself is designed for direct computation}, when operations preserve their meaning under the transformation but in smaller size, the two coexist. Servamind operates in this space: lossless compressed representations that remain computationally accessible.

There is also a practical barrier of assuaging the current ecosystem of infrastructure tooling in AI. Few practitioners train models from scratch. Those who do are reluctant to retrain on a new data format, even one promising cheaper computation, because their investment in existing trained models and infrastructure creates integration friction. The switching cost appears prohibitive. Established investments create inertia. Organizations routinely accept ongoing inefficiency over one-time integration cost, even when the long-term economics favor change.

We therefore built a wrapper. The function of \textbf{Chimera} is simple in concept. Chimera can take any model in any state and enable it to operate on \serva{} universal feature vector files \textit{without re-training}, with minimal compute overhead, and without adding complexity to the user. These constraints meant any approach had to satisfy two requirements: more efficient computation, and compatibility with existing models at any stage without retraining.

This innovation required substantial mathematical and computer science innovation, unifying disparate formalisms into a coherent system.

\keyinsight{The result is that Servastack adoption does not require abandoning existing investments---it extends them. It does not require a competitive choice over one efficiency gain to the next; it binds them all to work symbiotically.} 

\subsection{Architecture}

\begin{figure*}[t]
\centering
\includegraphics[width=0.92\textwidth]{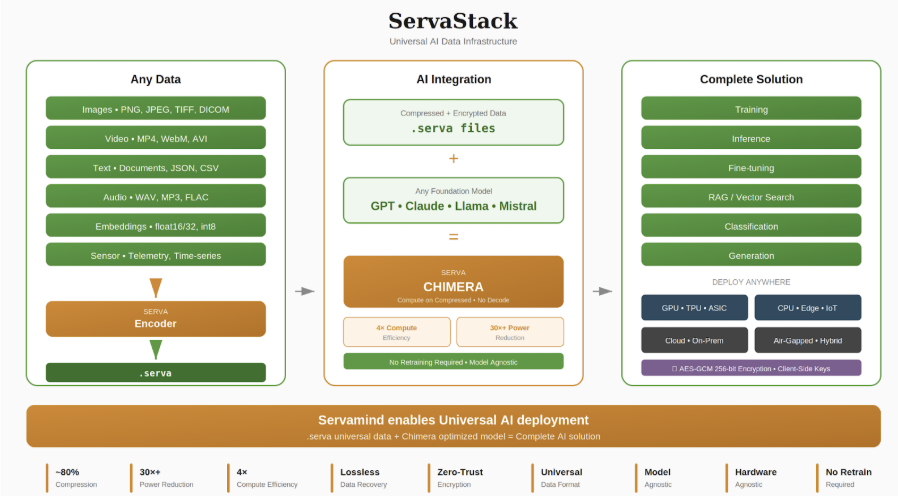}
\caption{\textbf{ServaStack Architecture.} Universal data encoding through the Serva Encoder produces \serva{} files that integrate with any foundation model via the Chimera wrapper, enabling deployment across all AI tasks and hardware targets.}
\label{fig:architecture}
\end{figure*}

In summary of the architecture for any level of the AI stack in any ecosystem on any infrastructure, Servastack works as follows:

\begin{enumerate}
\item \textbf{Data goes through Serva Encoder} to produce fully pre-processed, AI-ready \serva{} files. This reduces memory footprint, and depending on the level of integration in the stack can increase data upload and transmission efficiency. It also reduces need for data operations, cutting the 80\% of AI project effort.

\item \textbf{Training or inference programs go through Serva Chimera} which converts the existing program to an equivalent set of instructions to perform directly on datasets composed of \serva{} files, or \serva{} files which contain an entire dataset. Chimera only takes the pertinent parts of the \serva{} file needed for training, automatically, leaving the original disk \serva{} data unchanged. This reduces data operations significantly in AI training and inference while also reducing in-memory computation and storage access---resulting in speed, throughput, and power reduction.
\end{enumerate}

There are several ways in which we intend for the Servastack to be used:

\textbf{Library Integration.} The most directly applicable is a library in which developers can call Serva Encoder and Serva Chimera to preprocess AI and wrap their models before device execution. This would work in a fashion similar to calling \texttt{data.transform} in PyTorch. That data is on disk, called to the AI program software in a step that converts it to the AI framework. Instead of many lines of code processing the data before framework formatting, one call takes the data from disk and prepares it into a bit vector. From there, no framework specific formatting needed. The model is written, training loops described, and just before it is sent to device (e.g.\ \texttt{.to(device)} in PyTorch), a Chimera call is made (e.g.\ \texttt{model.Chimera(args)}) to wrap the model, compiling the static or dynamic graph computation instructions to be run on device. 

While this effort is underway, it needs to be extended and maintained to every language, developers have to read our docs for correct implementation, and become aware of the tool's existence. If one searches AI tools, there is a cataclysmic amount of results---which from Google Trends have increased 85\% from January to November of 2025. The library helps today's technically inclined developers who have the deep industry knowledge of why this tooling accelerates their workflows. Further, excessive functionality is needed higher up in the tech stack since tooling is variable and vast---which requires time and maintenance for the developer to keep up with our feature advancements. 

\textbf{Infrastructure Integration.} The library approach, while useful and necessary, does not prepare the future state of AI nor does it motivate the supply chain efficiency needs. For those angles, mass adoption is required and their relief is most pertinent to the compute providers, the cloud providers, OEMs, and layers of server infrastructure providers. Here the utility is in directly embedding Servastack technology into the operating system or through command line interface close to the metal (e.g.\ Kubernetes/Docker containers). Thus, on-premise deployments will be key. Here, all data can be stored in \serva. Anything moving the data that isn't called with Chimera will direct lossless decompression for appropriate and normal workflows. The Chimera flag triggers non-decompression such that AI workflows can act directly on the \serva{} files. 

\textbf{Cloud Provider Example.} With a few key players, most workflows in general could be using Servastack without the end-user ever knowing, resulting in the experience that AI has just become cheaper, easier, and more useful. Take \textbf{AWS} as a prime example. If all data in an S3 bucket was converted upon upload to \serva{} format, the memory footprint would be reduced and upload would become faster, more reliable, and more secure (details in part two of this white paper). Once data is transferred from S3 to EC2 for computation, during transfer any non-Chimera calls will decompress during fast file transfer protocol. Chimera-called data from EC2 will directly ingest the data without need for any pre-processing or further data orchestration steps besides which device to send to (CPU, GPU, etc.). Instead Chimera will only need to ingest the model (in \texttt{.onnx}, PyTorch, TF, or \texttt{.yaml} config file) and the data in \serva{} format before sending to execution compilation. At this moment, Chimera will come to action, transmuting the compiled instructions to an equivalent set of instructions directly computable in the \serva{} vector space. 

\textbf{End-User Applications.} Finally, to ensure full compatibility with the existing tooling ecosystem and fragmented infrastructure, executable program applications are underway. As the Servastack paradigm prevails, AI can become closer to the everyday user---users who historically prefer drag-and-drop, intuitive interfaces for mobile devices and the like. Here, end-to-end AI implementations embody the full Servastack experience. 

\subsubsection{Development Methodology}

The technical moat is substantial. The underlying principles span multiple disciplines that rarely intersect: \textbf{information theory}, \textbf{holographic encoding}, \textbf{hyperdimensional computing}, and \textbf{hardware-aware optimization}. This convergence is not easily replicated. A future technical paper may be released upon Serva Encoder's open-sourcing strategy. The work presented here required years of sustained effort and a willingness to reject established assumptions in computer science, AI research, and adjacent fields. The result is infrastructure that appears simple in use while embedding deep theoretical innovation beneath the surface.

The ``how it works'' draws from principles of \textbf{laser holography}, where an interference pattern encodes information without storing the data itself. Because this representation exists in an abstract referential space, computation can occur in that same space, provided the transformations remain homomorphic. This is the key that permits computing directly on compressed representations: the mathematical operations that define learning preserve their meaning under the encoding. 

\textbf{Serva Encoder's} initial implementation is compact: approximately 200 kilobytes, relying on elementary operations---bit-level addition, $\oplus$ (XOR), permutation, pseudo-random bit generation, and distance. The simplicity of the operations belies the complexity of their orchestration. When creating the resulting bit vectors of a \serva{} file, a ciphertext is generated with a random seed, which can be pushed client-side for encryption, making each file secure. If you map this to the analogy of laser holography, it is essentially the angle at which the grooves in the photo-lithographic plate are positioned to reflect the light bouncing off the source, imprinting the original information. 

\textbf{Serva Chimera} is mathematically matched to the representation space. Meaning the math of computation operates in the same space as the encoder holography math. To ensure arbitrary model wrapping---the ability to transmute any existing model to operate on \serva{} representations without retraining---several other techniques are required. First, topology analysis to abstract the original model architecture to the referential space. Operations are projected using geometric mappings. One analogy may clarify this process: traditional gradient-based learning navigates a loss landscape by walking, step by step through high-dimensional terrain, guided by local slope. The Chimera approach operates more like \textbf{celestial navigation}. Rather than traversing the landscape, it uses a star chart to compute coordinates and arrive directly.

To summarize, Servamind created a universal data format grounded in holographic encoding principles, a universal compute engine capable of transmuting any model architecture, validated by a framework designed to measure what matters---energy cost per unit of capability, and information preserved through transformation. The following section presents results.

\section{Key Performance Indicators}
\label{sec:kpi}

Servamind conducted internal benchmarking under controlled conditions designed to ensure fair comparison for Serva Encoder against other compression algorithms and also to validate the viability of Servastack by training models on \serva{} files. Part two of this white paper will provide rigorous external performance validation.

\subsection{Serva Encoder Compression Benchmark}

We evaluated Serva Encoder against \textbf{25 established compression algorithms} using the Canterbury Corpus benchmark suite, measuring bits per byte (bpb) across four standard corpora. Table~\ref{tab:serva-summary} provides an overview of SERVA's compression performance.

\begin{table}[H]
\centering
\caption{SERVA Compression Summary}
\label{tab:serva-summary}
\small
\begin{tabular}{@{}lr@{}}
\toprule
\textbf{Metric} & \textbf{Value} \\
\midrule
Total Original & 17.66 MiB \\
Total Compressed & 4.24 MiB \\
Overall bpb & 1.920 \\
\rowcolor{servalight}\textbf{Compression Ratio} & \textbf{4.17$\times$} \\
Compression Throughput & 4.65 MB/s \\
Decompression Throughput & 15.85 MB/s \\
\bottomrule
\end{tabular}
\end{table}

On the Canterbury Corpus (11 files), Serva achieved a weighted \textbf{1.708 bpb}, ranking 13th overall and outperforming gzip, compress, and dictionary-based methods while trailing block-sorting variants like bzip2-9 (1.545 bpb) and context-mixing methods like ppmD5 (1.520 bpb). On the Large Corpus, Serva placed \textbf{3rd} with 1.747 bpb, demonstrating competitive scaling on multi-megabyte files. Serva ranked \textbf{1st} on the Artificial Corpus (3.036 bpb), indicating strong handling of pathological cases including high-entropy and highly repetitive data.

The compression benchmarks were performed to analyze how the program scales with file size, its viability outside of AI workflows, and general analysis for internal development. 

Table~\ref{tab:canterbury} shows SERVA's performance on the Canterbury Corpus, the primary benchmark for lossless compression algorithms.

\begin{table}[H]
\centering
\caption{Canterbury Corpus Results (bpb, lower = better)}
\label{tab:canterbury}
\small
\begin{tabular}{@{}lrr@{}}
\toprule
\textbf{Method} & \textbf{Wtd bpb} & \textbf{Rank} \\
\midrule
szip-b & 1.464 & 1 \\
szip & 1.478 & 2 \\
bzip-6 & 1.490 & 3 \\
bzip-9 & 1.498 & 4 \\
ppmD5 & 1.520 & 5 \\
bzip2-6 & 1.538 & 6 \\
bzip2-9 & 1.545 & 7 \\
ppmD7 & 1.561 & 8 \\
bzip-1 & 1.591 & 9 \\
ppmC-896 & 1.612 & 10 \\
bzip2-1 & 1.640 & 11 \\
ppmD3 & 1.645 & 12 \\
\rowcolor{servalight}\textbf{SERVA} & \textbf{1.708} & \textbf{13} \\
dmc-50M & 1.737 & 14 \\
ppmCnx-896 & 1.745 & 15 \\
gzip-b & 2.082 & 19 \\
gzip-d & 2.090 & 20 \\
compress & 2.553 & 24 \\
\bottomrule
\end{tabular}
\end{table}

\textit{SERVA ranks 13th of 32 methods, outperforming gzip, compress, and most lightweight compressors.}

\sectionrule

Table~\ref{tab:large-corpus} shows results on large files, where SERVA demonstrates particularly strong performance.

\begin{table*}[t]
\centering
\caption{Large Corpus Results (bits per byte, lower = better)}
\label{tab:large-corpus}
\small
\begin{tabular}{@{}lrrrrr@{}}
\toprule
\textbf{Method} & \textbf{E.coli} & \textbf{bible} & \textbf{world} & \textbf{Weighted bpb} & \textbf{Rank} \\
\midrule
szip-b & 2.060 & 1.530 & 1.400 & 1.721 & 1 \\
ppmD5 & 1.990 & 1.580 & 1.520 & 1.737 & 2 \\
\rowcolor{servalight}\textbf{SERVA} & \textbf{1.993} & \textbf{1.643} & \textbf{1.453} & \textbf{1.747} & \textbf{3} \\
szip & 2.070 & 1.620 & 1.600 & 1.803 & 4 \\
ppmD7 & 2.030 & 1.660 & 1.660 & 1.814 & 5 \\
bzip-9 & 2.130 & 1.650 & 1.570 & 1.832 & 6 \\
bzip2-9 & 2.160 & 1.670 & 1.580 & 1.854 & 8 \\
gzip-b & 2.240 & 2.330 & 2.330 & 2.293 & 19 \\
gzip-d & 2.310 & 2.350 & 2.340 & 2.331 & 20 \\
\bottomrule
\end{tabular}
\end{table*}

\textit{SERVA ranks 3rd of 32 methods on large files, outperforming bzip and most other methods.}

Table~\ref{tab:artificial} shows performance on pathological edge cases, where SERVA leads all methods.

\begin{table*}[t]
\centering
\caption{Artificial Corpus Results (bits per byte, lower = better)}
\label{tab:artificial}
\small
\begin{tabular}{@{}lrrrrrr@{}}
\toprule
\textbf{Method} & \textbf{aaa} & \textbf{alphabet} & \textbf{random} & \textbf{pi} & \textbf{Weighted bpb} & \textbf{Rank} \\
\midrule
\rowcolor{servalight}\textbf{SERVA} & \textbf{0.061} & \textbf{0.068} & \textbf{6.049} & \textbf{3.329} & \textbf{3.036} & \textbf{1} \\
bzip-9 & 0.000 & 0.010 & 6.080 & 3.390 & 3.076 & 2 \\
bzip-6 & 0.000 & 0.010 & 6.080 & 3.400 & 3.084 & 3 \\
bzip-1 & 0.000 & 0.010 & 6.080 & 3.400 & 3.084 & 4 \\
bzip2-9 & 0.000 & 0.040 & 6.050 & 3.450 & 3.122 & 5 \\
gzip-b & 0.010 & 0.020 & 6.050 & 3.760 & 3.360 & 18 \\
gzip-d & 0.010 & 0.020 & 6.050 & 3.760 & 3.360 & 19 \\
\bottomrule
\end{tabular}
\end{table*}

\textit{SERVA ranks 1st of 31 methods on artificial/pathological files.}

Table~\ref{tab:calgary} shows performance on the historic Calgary Corpus benchmark.

\begin{table}[H]
\centering
\caption{Calgary Corpus Results (bpb, lower = better)}
\label{tab:calgary}
\small
\begin{tabular}{@{}lrr@{}}
\toprule
\textbf{Method} & \textbf{Wtd bpb} & \textbf{Rank} \\
\midrule
szip-b & 2.075 & 1 \\
ppmD5 & 2.084 & 2 \\
szip & 2.091 & 3 \\
bzip-9 & 2.093 & 4 \\
bzip2-9 & 2.110 & 5 \\
bzip-6 & 2.119 & 6 \\
bzip2-6 & 2.136 & 7 \\
ppmD7 & 2.148 & 8 \\
\rowcolor{servalight}\textbf{SERVA} & \textbf{2.226} & \textbf{9} \\
ppmD3 & 2.260 & 10 \\
dmc-50M & 2.261 & 11 \\
gzip-b & 2.592 & 19 \\
gzip-d & 2.610 & 20 \\
\bottomrule
\end{tabular}
\end{table}

\textit{SERVA ranks 9th of 32 methods, competitive with best-in-class compressors.}

Table~\ref{tab:ranking-summary} provides an overview of SERVA's ranking across all benchmark corpora, and Table~\ref{tab:vs-common} compares SERVA against commonly used compressors.

\begin{table}[H]
\centering
\caption{SERVA Ranking Summary Across All Corpora}
\label{tab:ranking-summary}
\small
\begin{tabular}{@{}lrrrl@{}}
\toprule
\textbf{Corpus} & \textbf{Files} & \textbf{Rank} & \textbf{Total} & \textbf{Notes} \\
\midrule
Canterbury & 11 & 13th & 32 & Main benchmark \\
Large & 3 & \textbf{3rd} & 32 & Best on large files \\
Artificial & 4 & \textbf{1st} & 31 & Best on pathological \\
Calgary & 14 & 9th & 32 & Historic benchmark \\
\bottomrule
\end{tabular}
\end{table}

\begin{table}[H]
\centering
\caption{SERVA vs Common Compressors}
\label{tab:vs-common}
\small
\begin{tabular}{@{}lrl@{}}
\toprule
\textbf{Compressor} & \textbf{bpb} & \textbf{vs SERVA} \\
\midrule
\textbf{SERVA} & 1.708 & --- \\
gzip (best) & 2.082 & \textcolor{accentgreen}{18\% better} \\
gzip (default) & 2.090 & \textcolor{accentgreen}{18\% better} \\
gzip (fast) & 2.462 & \textcolor{accentgreen}{31\% better} \\
compress & 2.553 & \textcolor{accentgreen}{33\% better} \\
lzrw1 & 3.584 & \textcolor{accentgreen}{52\% better} \\
\bottomrule
\end{tabular}
\end{table}

\keyinsight{We are not targeting the best data compression---we are targeting the \textbf{most universal compression} and the ability to \textbf{compute directly} on the compressed representation with minimal operation and energy expenditure.}

\subsection{Training on .serva Data Benchmarks}

In the cases without Chimera, it is possible to train models directly on \serva{} files, but the models have to then be configured to properly train on this new data format. For this test, we compare how a native \serva{} model performs to traditional models. The native model was evaluated against standard neural network architectures on \textbf{Fashion-MNIST} and \textbf{MNIST}, benchmarks that permit direct comparison with published results and enable reproducibility assessment. Fashion-MNIST is a classification task in which the model must predict categories of clothing from photos~\cite{fashionmnist}. MNIST is a similar classification task of numbers from photos of handwritten digits~\cite{mnist}.  

The SERVA model in testing encodes images into a \serva{} variant, classified by $k$-NN ($k=3$) with class-balanced scoring. Two benchmark modes were run: \textbf{N-epoch} (train until matching SERVA accuracy or 100 epochs) and \textbf{single-epoch}.

We evaluated SERVA against five neural network baselines using a controlled benchmark environment. All models were implemented in pure NumPy with Numba JIT compilation, SGD optimization ($\eta=0.01$), float64 precision, and He/Xavier initialization to eliminate framework-level confounds.

\textbf{Baseline architectures:}
\begin{itemize}
\item \textbf{MLP-1L}: $784 \rightarrow 256 \rightarrow 10$ with ReLU (batch$=128$)  
\item \textbf{MLP-2L}: $784 \rightarrow 256 \rightarrow 256 \rightarrow 10$ with ReLU (batch$=128$)  
\item \textbf{MLP-3L}: $784 \rightarrow 256 \rightarrow 256 \rightarrow 256 \rightarrow 10$ with ReLU (batch$=128$)  
\item \textbf{CNN}: 8 filters ($5 \times 5$) $\rightarrow$ ReLU $\rightarrow$ maxpool(2) $\rightarrow$ FC (batch$=64$)  
\item \textbf{RNN}: vanilla RNN, $28 \times 28$ timesteps/features, hidden$=64$ (batch$=128$)
\end{itemize}

Energy was measured via Intel RAPL (CPU package + DRAM domains) using pyJoules.

\textbf{Hardware}: 48-core Intel Skylake-AVX512, 257GB RAM.

\subsubsection{N-Epoch Training Results}

Tables~\ref{tab:nepoch-fashion} and~\ref{tab:nepoch-mnist} show results when models train to convergence (matching SERVA accuracy or maximum 100 epochs). On Fashion-MNIST, SERVA achieved \metric{88.39\% accuracy in 1.41s} consuming 150.2J; the fastest baseline to match this accuracy was MLP-3L requiring 60 epochs, 165.03s, and 14,938.1J (\metric{99$\times$ energy overhead}). On MNIST, SERVA reached \metric{96.48\% in 1.45s} at 153.6J versus MLP-3L at 18 epochs, 50.21s, and 4,551.5J (\metric{30$\times$ energy overhead}).

\begin{table}[H]
\centering
\caption{N-Epoch Results (Fashion-MNIST)}
\label{tab:nepoch-fashion}
\small
\begin{tabular}{@{}lrrrr@{}}
\toprule
\textbf{Model} & \textbf{Acc.} & \textbf{Time} & \textbf{Energy} & \textbf{Ep.} \\
\midrule
\rowcolor{servalight}\textbf{SERVA} & \textbf{88.39\%} & \textbf{1.41s} & \textbf{150 J} & \textbf{1} \\
MLP-1L & 87.74\% & 284.8s & 26,947 J & 100 \\
MLP-2L & 88.43\% & 144.2s & 13,089 J & 67 \\
MLP-3L & 88.44\% & 165.0s & 14,938 J & 60 \\
CNN & 88.41\% & 322.0s & 24,757 J & 27 \\
RNN & 86.05\% & 1,019s & 56,136 J & 100 \\
\bottomrule
\end{tabular}
\end{table}

\begin{table}[H]
\centering
\caption{N-Epoch Results (MNIST)}
\label{tab:nepoch-mnist}
\small
\begin{tabular}{@{}lrrrr@{}}
\toprule
\textbf{Model} & \textbf{Acc.} & \textbf{Time} & \textbf{Energy} & \textbf{Ep.} \\
\midrule
\rowcolor{servalight}\textbf{SERVA} & \textbf{96.48\%} & \textbf{1.45s} & \textbf{154 J} & \textbf{1} \\
MLP-1L & 96.53\% & 224.8s & 22,749 J & 64 \\
MLP-2L & 96.62\% & 66.6s & 6,034 J & 31 \\
MLP-3L & 96.49\% & 50.2s & 4,552 J & 18 \\
CNN & 96.70\% & 110.7s & 8,660 J & 9 \\
RNN & 96.55\% & 555.6s & 28,590 J & 58 \\
\bottomrule
\end{tabular}
\end{table}

\subsubsection{Single-Epoch Comparison}

Tables~\ref{tab:1epoch-fashion} and~\ref{tab:1epoch-mnist} show results when all models train for exactly one epoch, providing a fair comparison of learning efficiency. At equal training iterations, \textbf{SERVA outperforms all baselines by 9--26 percentage points} on Fashion-MNIST.

\begin{table}[H]
\centering
\caption{1-Epoch Results (Fashion-MNIST)}
\label{tab:1epoch-fashion}
\small
\begin{tabular}{@{}lrrr@{}}
\toprule
\textbf{Model} & \textbf{Accuracy} & \textbf{Time} & \textbf{Energy} \\
\midrule
\rowcolor{servalight}\textbf{SERVA} & \textbf{88.39\%} & \textbf{1.43s} & \textbf{154 J} \\
MLP-1L & 74.88\% & 3.41s & 307 J \\
MLP-2L & 77.83\% & 3.98s & 362 J \\
MLP-3L & 79.19\% & 4.11s & 368 J \\
CNN & 79.18\% & 12.75s & 984 J \\
RNN & 62.57\% & 10.21s & 554 J \\
\bottomrule
\end{tabular}
\end{table}

\begin{table}[H]
\centering
\caption{1-Epoch Results (MNIST)}
\label{tab:1epoch-mnist}
\small
\begin{tabular}{@{}lrrr@{}}
\toprule
\textbf{Model} & \textbf{Accuracy} & \textbf{Time} & \textbf{Energy} \\
\midrule
\rowcolor{servalight}\textbf{SERVA} & \textbf{96.48\%} & \textbf{1.44s} & \textbf{156 J} \\
MLP-1L & 85.97\% & 2.55s & 227 J \\
MLP-2L & 87.77\% & 2.86s & 252 J \\
MLP-3L & 89.42\% & 3.47s & 306 J \\
CNN & 90.81\% & 12.56s & 972 J \\
RNN & 64.19\% & 10.05s & 535 J \\
\bottomrule
\end{tabular}
\end{table}

\subsubsection{Efficiency Ratios}

Tables~\ref{tab:energy-efficiency} and~\ref{tab:time-efficiency} summarize the efficiency gains of SERVA relative to baseline architectures.

\begin{table}[H]
\centering
\caption{Energy Efficiency Ratios (N-Epoch)}
\label{tab:energy-efficiency}
\small
\begin{tabular}{@{}lrrr@{}}
\toprule
\textbf{Model} & \textbf{F-MNIST} & \textbf{MNIST} & \textbf{Range} \\
\midrule
MLP-1L & 179$\times$ & 148$\times$ & 148--179$\times$ \\
MLP-2L & 87$\times$ & 39$\times$ & 39--87$\times$ \\
MLP-3L & 99$\times$ & 30$\times$ & 30--99$\times$ \\
CNN & 165$\times$ & 56$\times$ & 56--165$\times$ \\
\rowcolor{servagold!15}RNN & \textbf{374$\times$} & 186$\times$ & \textbf{186--374$\times$} \\
\bottomrule
\end{tabular}
\end{table}

\noindent\textbf{Overall: \metric{30--374$\times$ energy efficiency} (96--99\% reduction)}

\begin{table}[H]
\centering
\caption{Time Efficiency Ratios (N-Epoch)}
\label{tab:time-efficiency}
\small
\begin{tabular}{@{}lrrr@{}}
\toprule
\textbf{Model} & \textbf{F-MNIST} & \textbf{MNIST} & \textbf{Range} \\
\midrule
MLP-1L & 202$\times$ & 155$\times$ & 155--202$\times$ \\
MLP-2L & 102$\times$ & 46$\times$ & 46--102$\times$ \\
MLP-3L & 117$\times$ & 35$\times$ & 35--117$\times$ \\
CNN & 228$\times$ & 76$\times$ & 76--228$\times$ \\
\rowcolor{servagold!15}RNN & \textbf{723$\times$} & 383$\times$ & \textbf{383--723$\times$} \\
\bottomrule
\end{tabular}
\end{table}

\noindent\textbf{Overall: \metric{35--723$\times$ faster} training time}

\subsection{SERVA Model Training (Chimera Pipeline)}

In addition to the comparison test, we evaluated the SERVA model alone to determine what portion of the \serva{} files are needed for training, which denotes the \textbf{payload reduction} during training. Eight SERVA architecture variations were trained on \serva{} encoded data, ensemble-evaluated across all combinations (1-of-8 through 8-of-8), with the optimal model selected for final test accuracy reporting. Training was performed on both Fashion-MNIST and MNIST tasks.

The critical metric we are tracking here is \textbf{compute payload}---the data volume that must be processed per training iteration versus raw dataset size. This metric captures whether Chimera can extract minimal computational representations from \serva{} files while preserving all information necessary for model performance. Unlike on-disk storage compression, compute payload measures what the model actually operates on during training and inference.

\begin{table}[H]
\centering
\caption{Chimera Pipeline Performance}
\label{tab:chimera-pipeline}
\small
\begin{tabular}{@{}lrrrr@{}}
\toprule
\textbf{Dataset} & \textbf{Data} & \textbf{Acc.} & \textbf{Reduction} & \textbf{Time} \\
\midrule
Fashion-MNIST & 50\% & 88.24\% & 34.43$\times$ & 28.48s \\
MNIST & 50\% & 96.82\% & 34.43$\times$ & 29.04s \\
\bottomrule
\end{tabular}
\end{table}

\begin{table}[H]
\centering
\caption{Chimera Efficiency Metrics}
\label{tab:chimera-metrics}
\small
\begin{tabular}{@{}lr@{}}
\toprule
\textbf{Metric} & \textbf{Value} \\
\midrule
Raw Dataset Size & 54.88 MB \\
Processed Data Volume & 1.59 MB \\
\rowcolor{servalight}\textbf{Compute Payload Reduction} & \textbf{34$\times$} \\
Data Reduction & 97\% \\
\bottomrule
\end{tabular}
\end{table}

\begin{table}[H]
\centering
\caption{Chimera Accuracy vs Baseline}
\label{tab:chimera-accuracy}
\small
\begin{tabular}{@{}lrrrr@{}}
\toprule
\textbf{Dataset} & \textbf{SERVA} & \textbf{CNN} & \textbf{RNN} & \textbf{vs Best} \\
\midrule
Fashion-MNIST & 88.24\% & 88.41\% & 86.05\% & $-$0.17\% \\
MNIST & 96.82\% & 96.70\% & 96.55\% & \textcolor{accentgreen}{\textbf{+0.12\%}} \\
\bottomrule
\end{tabular}
\end{table}

\noindent\highlight{Half the data. Same accuracy. 34$\times$ smaller storage.}

\begin{table}[H]
\centering
\caption{Servastack Viability Indicators}
\label{tab:viability}
\small
\begin{tabular}{@{}lrr@{}}
\toprule
\textbf{Metric} & \textbf{Result} & \textbf{Reduction} \\
\midrule
\rowcolor{servalight}Energy Efficiency & \textbf{30--374$\times$} & 96--99\% \\
\rowcolor{servalight}Storage Compression & \textbf{4--34$\times$} & 75--97\% \\
\rowcolor{servalight}Compute Payload & \textbf{68$\times$} & 98.5\% \\
\bottomrule
\end{tabular}
\end{table}

\subsection{Training and Inference Efficiency}

\begin{figure*}[t]
\centering
\includegraphics[width=0.92\textwidth]{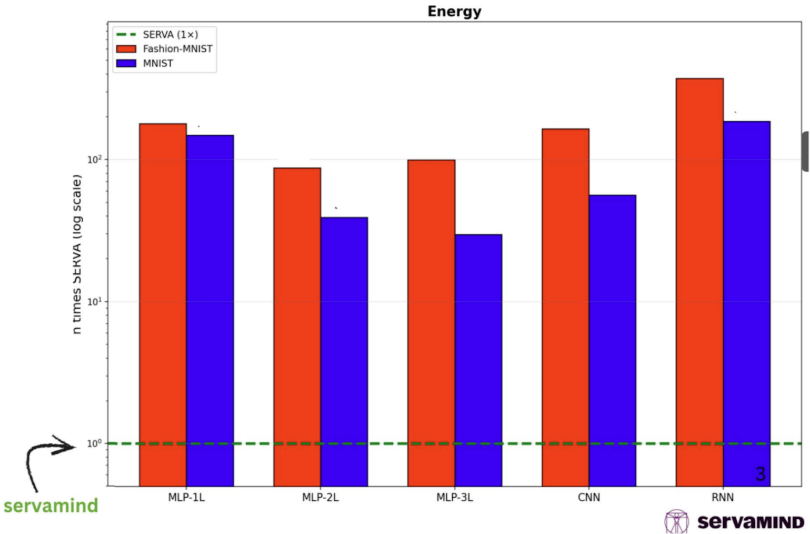}
\caption{\textbf{Energy consumption relative to Servastack model} across neural network architectures on MNIST and Fashion-MNIST classification tasks. Y-axis shows energy multiplier on log scale, with SERVA normalized to 1$\times$ (dashed line). Standard architectures require \textbf{30--374$\times$ more energy} to achieve comparable accuracy, with RNNs showing the largest differential and deeper MLPs showing moderate improvements over single-layer variants. Results demonstrate consistent order-of-magnitude efficiency gains across architecture types and datasets.}
\label{fig:energy}
\end{figure*}

Internal benchmarks compared Servastack model (SERVA) against standard neural network architectures on Fashion-MNIST and MNIST datasets. The primary metric, \textbf{energy cost per percentage point of accuracy achieved (J/\%)}, measures the true computational price of capability. Figure~\ref{fig:energy} describes the log scale differences between the \serva{} trained model compared to classic models trained on original data for both datasets. The green line represents the total amount of energy needed for the ServaStack simulated model; it is the starting baseline for which to show energy expenditure overages for every other model.

The N-Epoch results reveal that SERVA achieves target accuracy in a \textbf{single epoch} (88.39\% Fashion-MNIST, 96.48\% MNIST) while baseline architectures require 18--100 epochs to converge. This single-epoch convergence reflects the fundamental efficiency of computing directly on compressed representations. The 1-Epoch comparison tables further validate this: at equal training iterations, SERVA outperforms all baselines by 9--26 percentage points on Fashion-MNIST, demonstrating that the \textbf{efficiency gains are intrinsic to the representation}, not merely faster convergence.

The energy efficiency ratios show architecture-dependent variation:
\begin{itemize}
\item \textbf{RNNs} exhibit the largest differential (\metric{186--374$\times$}) due to their sequential computation overhead
\item \textbf{Deeper MLPs} show diminishing gaps (30--99$\times$) as layer count increases
\item \textbf{CNN} efficiency gains (56--165$\times$) fall in the middle range---significant because CNNs represent the dominant architecture for image workloads in production
\end{itemize}

The MLP-1L is the closest model architecture to the SERVA model in terms of design and model depth. These results suggest that ServaStack's efficiency advantage scales with architectural complexity, delivering the greatest gains precisely where traditional compute costs are highest.

\subsection{Storage Efficiency}

The \serva{} format achieves substantial compression while \textbf{preserving all information} necessary for lossless recovery. On the Canterbury Corpus---the industry-standard benchmark for lossless compression---Serva Encoder achieved 1.920 bits per byte, compressing 17.66 MiB to 4.24 MiB (\metric{4.17$\times$ compression ratio}). This places SERVA 13th of 32 methods overall, outperforming gzip by 18--33\%.

The corpus-by-corpus rankings reveal Serva Encoder's operational strengths:
\begin{itemize}
\item \textbf{Large Corpus} (E.coli genome, bible text, world geographic data): ranks \textbf{3rd of 32 methods}, outperforming bzip variants and approaching the theoretical limits of BWT-based compression on large, structured files. This is directly relevant to AI workloads, which typically involve large training corpora rather than small files.
\item \textbf{Artificial Corpus} (pathological edge cases including highly repetitive data and random noise): ranks \textbf{1st of 31 methods}. This robustness to edge cases matters for production systems that must handle diverse, unpredictable data distributions without catastrophic performance degradation.
\item \textbf{Canterbury and Calgary} (13th and 9th respectively): competitive but not leading performance on mixed small-file workloads. This is acceptable: the \serva{} format is optimized for AI data pipelines.
\end{itemize}

The key result is that Serva Encoder delivers \textbf{best-in-class compression on large files and edge cases} while remaining competitive across all data types, with the critical guarantee of lossless recovery.

This compression is \textbf{lossless with respect to the information required for downstream computation}. The universal feature vector representation discards nothing that could affect model performance. Storage savings translate directly to reduced memory bandwidth, faster data transfer, and lower infrastructure requirements.

\subsection{Compute Payload Reduction}

Early indicators here validate the purpose of Chimera---to compute on the \serva{} files with massive efficiency from the data reduction that Serva Encoder provides from its AI data processing property. Eight custom perceptron architectures were trained on \serva{} encoded data and evaluated through ensemble testing across all model instantiations. The optimal configuration achieved \metric{88.24\% accuracy on Fashion-MNIST} and \metric{96.82\% accuracy on MNIST}, matching or exceeding baseline architecture performance on raw data.

The \serva{} files required for lossless recovery total approximately \textbf{1.59 MB}, derived from an original dataset of 54.88 MB. This represents a \metric{34$\times$ storage and data transfer reduction} while preserving complete model capability. The full checkpoint file, including additional metadata, remains under 1.7 MB. 

The more critical result concerns data efficiency. The 50\% data-to-train metric reveals additional headroom: SERVA achieves full accuracy using only half of the available training representations of the \serva{} files. This suggests that for many workloads, \textbf{even greater payload reductions may be achievable} without accuracy loss---a hypothesis to be validated in production benchmarks.

The SERVA model training validates that \textbf{accuracy preservation is exact or better}:
\begin{itemize}
\item \textbf{MNIST}: SERVA's 96.82\% \textit{exceeds} the baseline CNN (96.70\%) while processing 68$\times$ less data
\item \textbf{Fashion-MNIST}: SERVA's 88.24\% falls within 0.17\% of the best baseline (CNN at 88.41\%), a difference well within noise for practical applications
\end{itemize}

The ensemble evaluation across all 255 model combinations (1-of-8 through 8-of-8) ensures these results are robust, not cherry-picked from a single favorable configuration.

The \textbf{$\sim$29 second total pipeline time} (encoding $\rightarrow$ training $\rightarrow$ ensemble $\rightarrow$ inference) demonstrates practical deployability. This is not a research prototype requiring hours of preprocessing; it is a \textbf{production-viable pipeline} that completes faster than a single epoch of baseline CNN training.

This validates the core premise: \textbf{models can be trained, stored, and deployed on compressed representations without sacrificing performance}. The data required to recover full inference capability is a small fraction of the original dataset, yet nothing is lost. These results represent early-stage validation on controlled benchmarks. Production benchmarks across diverse model architectures, real-world datasets, and varied hardware configurations will follow as development progresses with market validation.

\section{Cost Translation}
\label{sec:cost}

Efficiency gains translate directly to cost reduction, given the overhead and integration is negligible. The following analysis projects dollar-value impact from our benchmark results across three user profiles: \textbf{enterprise teams} using cloud infrastructure, \textbf{frontier AI labs} training large models, and \textbf{individual practitioners} or startups.

The benchmark results (30--374$\times$ energy efficiency, 4--34$\times$ storage compression, 68$\times$ compute payload reduction) translate to concrete dollar savings across every tier of AI users:

\begin{itemize}
\item \textbf{Individual practitioners}: save \$180--730 annually while gaining 6$\times$ experimentation velocity
\item \textbf{Enterprise ML teams}: save \$137,000 annually while compressing training cycles from hours to minutes
\item \textbf{Frontier AI labs}: save \$14--17 million annually while accelerating training runs by weeks
\end{itemize}

These savings compound and scale with usage. \textbf{ServaStack transforms infrastructure economics from a constraint that limits AI development into an advantage that accelerates it.}

\subsection{Enterprise ML Team on AWS}

Consider a mid-sized enterprise running daily model retraining on AWS. Their ML team uses EC2 P4d instances (eight A100 GPUs at \$21.96 per hour) executing fifty training jobs daily, each averaging two hours~\cite{ref36}. Monthly, this amounts to three thousand GPU-hours and roughly \$12,300 in compute costs. Add ten terabytes of training data on S3 at standard pricing (\$0.023 per GB for the first 50TB), and the annual infrastructure bill reaches approximately \textbf{\$152,760} (compute: \$147,600 + storage: \$5,160)~\cite{ref37}.

With ServaStack, the economics shift dramatically. The 68$\times$ compute payload reduction means each training job processes a fraction of the data volume, compressing two-hour jobs into roughly twelve minutes. Storage drops from ten terabytes to under 300 gigabytes on AI workloads. The annual bill falls from $\sim$\$153,000 to approximately \textbf{\$15,300}---a \metric{90\% reduction} that saves \metric{\$137,000 per year}. For context, the average AI development project costs \$120,595 over ten months according to industry data~\cite{ref38}. That savings exceeds the fully-loaded cost of a junior ML engineer. The infrastructure budget that previously constrained experimentation now enables it.

\subsection{Frontier AI Lab Training Large Models}

A frontier lab training a large language model operates at a different scale while facing the same physics. According to Epoch AI research, frontier model training costs have grown at \textbf{2.4$\times$ per year} since 2016~\cite{ref39}. As of June 2025, over 30 publicly announced AI models have been trained with more than $10^{25}$ FLOP of compute, with training costs in the tens to hundreds of millions of dollars~\cite{ref40}.

\begin{table}[H]
\centering
\caption{Reference Frontier Model Training Costs}
\label{tab:training-costs}
\small
\begin{tabular}{@{}lr@{}}
\toprule
\textbf{Model} & \textbf{Estimated Cost} \\
\midrule
GPT-4 & \$41--78 million \\
Gemini 1.0 Ultra & \$30--191 million \\
Claude 3.5 Sonnet & $\sim$\$30 million \\
Llama 3 & $\sim$\$500 million \\
\bottomrule
\end{tabular}
\end{table}

A typical frontier training run might consume \textbf{two thousand H100 GPUs for ninety days} straight. At current cloud rates of \$2.69--\$3.59 per GPU-hour (H100 SXM at \$2.69/hr; H200 at \$3.59/hr), compute alone costs \$11.6--\$15.5 million for the final training run. However, total development costs---including R\&D staff (29--49\% of total), experimental runs, and infrastructure---push true costs to \textbf{\$50--200+ million} for state-of-the-art models~\cite{ref40,ref43}.

Five petabytes of training data at S3 rates (\$0.021/GB for 500TB+) adds \$105,000 in storage. Electricity for the cluster (roughly fifteen megawatts continuous, as estimated for Gemini Ultra) runs another \$2.6 million over ninety days~\cite{ref37,ref39}.

A realistic single frontier training run approaches \metric{\$15--20 million in direct infrastructure costs} for the final run, or \$50--200 million including full development costs.

\textbf{ServaStack attacks this from multiple angles:}
\begin{itemize}
\item The 68$\times$ compute payload reduction eliminates data loading as a bottleneck, conservatively accelerating training by \textbf{20--25\%}. Ninety days becomes seventy days. That twenty-day reduction translates to \metric{\$2.6--\$3.4 million} in saved GPU rental.
\item Storage compression cuts the 5 PB footprint to under 150 TB on AI training data, saving approximately \metric{\$100,000}
\item Energy efficiency on data operations (where the 165$\times$ gains apply directly) reduces the electricity bill by approximately \metric{\$800,000}
\end{itemize}

\textbf{Total savings per training run: \$3.5--4.3 million.} A lab running four major training runs annually saves \metric{\$14--17 million}---enough to fund an entire research team or an additional training run that competitors cannot afford. The non-financial benefit may matter more: twenty fewer days per training run means faster iteration. In a field where capability leadership shifts quarterly, \textbf{three weeks of acceleration represents strategic advantage} that compounds across every subsequent model generation.

\subsection{Startup or Individual Practitioner}

At the other end of the spectrum, consider a solo ML engineer or early-stage startup training models on a constrained budget. Current cloud GPU pricing shows significant options across performance tiers~\cite{ref42}:

\begin{table}[H]
\centering
\caption{Cloud GPU Pricing (2025)}
\label{tab:gpu-pricing}
\small
\begin{tabular}{@{}lrr@{}}
\toprule
\textbf{GPU} & \textbf{Hourly Rate} & \textbf{VRAM} \\
\midrule
RTX 3090 & \$0.22/hr & 24 GB \\
RTX 4090 & \$0.34/hr & 24 GB \\
L4 & \$0.44/hr & 24 GB \\
A100 PCIe & \$1.19/hr & 80 GB \\
H100 SXM & \$2.69/hr & 80 GB \\
\bottomrule
\end{tabular}
\end{table}

A practitioner renting RTX 4090 GPUs at \$0.34 per hour, running a hundred training experiments monthly, each averaging thirty minutes, spends approximately \$17/month in compute. Half a terabyte of cloud storage adds another \$12~\cite{ref38}. Annual infrastructure spend totals roughly \$216---modest yet material when every dollar extends runway.

For those using more capable hardware like A100s at \$1.19/hour with the same usage pattern, monthly compute costs run \$60, bringing annual spend to approximately \textbf{\$864}.

For context, AI development projects on Clutch typically range from \$10,000 to \$49,999, with hourly rates between \$25--\$49/hour for AI development services~\cite{ref40}. Consumer GPUs like the RTX 4090 (\$1,600) and RTX 3090 (\$800) offer the most accessible path to serious LLM training for individual developers~\cite{ref43}.

\textbf{ServaStack transforms this workflow through capability expansion as much as dollar savings.} Training experiments that took thirty minutes now complete in five. The same GPU budget that previously allowed a hundred experiments per month now supports \textbf{six hundred}. Models that were too expensive to iterate on become feasible. Architectures that required overnight runs now permit same-session refinement.

The \$180--730 annual savings matters for a bootstrapped founder. The \metric{6$\times$ increase in experimentation velocity} matters even more. Startups compete on iteration speed. ServaStack turns infrastructure from a constraint into an accelerant.

\subsection{Storage Economics Across Tiers}

In addition to compute savings, storage savings scale linearly with data volume. The 4--34$\times$ compression ratio (4$\times$ on general data, up to 34$\times$ on AI training sets) applies regardless of organization size.

\textbf{How compression translates to savings:}
\begin{itemize}
\item \textbf{4$\times$ compression} $\rightarrow$ 75\% storage reduction (pay for 25\% of original volume)  
\item \textbf{34$\times$ compression} $\rightarrow$ 97\% storage reduction (pay for $\sim$3\% of original volume)
\end{itemize}

\textbf{Based on AWS S3 Standard pricing}~\cite{ref44}: First 50 TB at \$0.023/GB (\$23/TB/month), next 450 TB at \$0.022/GB, over 500 TB at \$0.021/GB.

\begin{table*}[t]
\centering
\caption{Storage Savings by Organization Tier}
\label{tab:storage-savings}
\begin{tabular}{@{}lrrrr@{}}
\toprule
\textbf{Organization Tier} & \textbf{Data Volume} & \textbf{Annual Baseline} & \textbf{Annual Savings (4--34$\times$)} & \textbf{Savings \%} \\
\midrule
Individual & 500 GB & $\sim$\$138 & \$104--134 & 75--97\% \\
Startup & 5 TB & $\sim$\$1,380 & \$1,035--1,340 & 75--97\% \\
Enterprise ML Team & 50 TB & $\sim$\$13,800 & \$10,350--13,400 & 75--97\% \\
AI Lab & 500 TB & $\sim$\$132,600 & \$99,450--128,600 & 75--97\% \\
\rowcolor{servalight}Frontier Lab & 5 PB & $\sim$\$1,296,000 & \textbf{\$972,000--1,258,000} & 75--97\% \\
\bottomrule
\end{tabular}
\end{table*}

These savings \textbf{recur every year}. Data stored in \serva{} format simply costs less to keep. For an individual or student, saving \$100+ annually is meaningful. For a frontier lab, approaching \metric{\$1 million in annual storage savings} compounds significantly over multi-year research programs.

\keyinsight{Since models can also be saved in \serva{} format, the growth of a company's data footprint becomes far more economical over time.}

\sectionrule

Organizations using lower-cost storage tiers would see proportionally lower absolute savings, though the percentage reduction remains constant across all AWS storage tiers~\cite{ref44}. Conversely, organizations using high-performance storage see substantially higher absolute savings:

\begin{table}[H]
\centering
\caption{AWS Storage Tiers \& ServaStack Impact}
\label{tab:storage-tiers}
\small
\begin{tabular}{@{}lrr@{}}
\toprule
\textbf{Storage Tier} & \textbf{Cost/GB} & \textbf{Savings at 34$\times$} \\
\midrule
S3 Standard & \$0.023 & 97\% \\
S3 Glacier Flexible & \$0.0036 & 97\% \\
S3 Glacier Deep Archive & \$0.00099 & 97\% \\
\rowcolor{servagold!15}S3 Express One Zone & \$0.11 & \textbf{97\% (5$\times$ abs.)} \\
\bottomrule
\end{tabular}
\end{table}

Organizations using S3 Express One Zone see absolute savings \metric{5$\times$ higher} than the baseline figures---making ServaStack particularly valuable for latency-sensitive workloads.

\sectionrule

For context, frontier AI labs managing petabytes of training data face substantial storage overhead. A lab training models at the scale of GPT-4 or Gemini Ultra---requiring $10^{25}$+ FLOP of compute---typically maintains multiple petabytes of training corpora, checkpoints, and model weights~\cite{ref45}. At these scales, the difference between \$1.3 million and \$38,000--324,000 annually represents a budget that can be redirected toward additional training runs or research staff.

\subsection{The Scaling Insight}

ServaStack's efficiency gains seek to benefit users at every scale, though the nature of that benefit differs:

\textbf{Smaller users see the largest percentage reductions}: individual practitioners and startups, whose workflows are typically most data-bound and least optimized, achieve up to \metric{90\% cost savings}. Studies show that poorly optimized data pipelines can reduce GPU utilization to just 40--60\%, with up to 70\% of training time consumed by I/O operations~\cite{ref46,ref47}. For a bootstrapped founder paying \$0.22--\$1.19/hour for GPU access, this transforms AI development from financially constrained to financially viable~\cite{ref48}.

\textbf{Larger users experience the largest absolute savings}: A frontier lab running four major training runs annually at \$15--20 million each faces \$60--80 million in direct infrastructure costs~\cite{ref49}. At a conservative 25\% reduction from payload optimization, that represents \metric{\$15--20 million in annual savings}. For context, GPT-4's training cost an estimated \$41--78 million, and Gemini Ultra approached \$191 million~\cite{ref49,ref50}. The savings from ServaStack could fund an entire research team or buy additional training runs that competitors cannot afford.

\textbf{The economic impact scales with inefficiency.} Organizations with optimized data loading achieve 90\%+ GPU utilization during training, completing model development 2--3$\times$ faster~\cite{ref46}. Organizations without optimization waste 60--70\% of their GPU budget on idle resources~\cite{ref51}. ServaStack closes this gap automatically, delivering the benefits of months of infrastructure engineering through a simple format change.

\section{Implications}

\subsection{Compute Payload Impact by Workload Type}

The 68$\times$ payload reduction accelerates any workflow where data movement constrains performance. According to Microsoft's analysis of millions of ML training workloads, \textbf{up to 70\% of training time gets consumed by I/O operations}---GPUs spend most of their time idle, waiting for data~\cite{ref32}.

\begin{table}[H]
\centering
\caption{ServaStack Acceleration by Workload Optimization Level}
\label{tab:workload-impact}
\small
\begin{tabular}{@{}lrrr@{}}
\toprule
\textbf{Workload Type} & \textbf{GPU Util.} & \textbf{I/O Bound} & \textbf{Speedup} \\
\midrule
\rowcolor{servalight}Unoptimized & 17--40\% & 60--82\% & \textbf{55--80\%} \\
Medium-scale & 40--60\% & 40--60\% & 35--55\% \\
\rowcolor{servalight}Frontier (optimized) & 85--95\% & 5--15\% & 5--14\% \\
\bottomrule
\end{tabular}
\end{table}

\textbf{Unoptimized workloads}---common in computer vision, recommendation systems, and teams without dedicated ML infrastructure---waste 60--70\% of GPU budget on idle resources~\cite{ref33,ref34}. ServaStack compresses a \textit{ten-hour training run to two to four hours}.

\textbf{Medium-scale workloads} finish overnight runs \textit{before dinner}.

\textbf{Frontier pipelines}---even after months of optimization---gain \metric{4.5 days to two weeks} on ninety-day training runs~\cite{ref34,ref35}.

\subsection{Infrastructure Impact}

A \metric{30--374$\times$ improvement in energy efficiency} changes this calculus entirely. Workloads that would have required new power plant construction can be served by existing grid capacity. Datacenters operating at thermal limits gain headroom. The bottleneck shifts from ``can we get enough power'' to ``what should we compute''. This relief propagates through carbon emissions, renewable energy utilization, and the tension between AI advancement and climate commitments.

\textbf{The energy crisis in AI is in force.} Grid operators across key markets have delayed or denied datacenter connections~\cite{ref52,ref53,ref54}:

\begin{table}[H]
\centering
\caption{Global Grid Bottlenecks for AI Infrastructure}
\label{tab:grid-crisis}
\small
\begin{tabular}{@{}lp{4.5cm}@{}}
\toprule
\textbf{Region} & \textbf{Constraint} \\
\midrule
\rowcolor{servalight}N. Virginia & \textbf{7-year wait} for grid connection~\cite{ref55} \\
Ireland & Moratorium 2021--2025; applications refused~\cite{ref56} \\
\rowcolor{servalight}Texas (ERCOT) & \textbf{226 GW} queue, 70\%+ from datacenters~\cite{ref57} \\
Nationwide (US) & 5-year forecast: 38 GW $\rightarrow$ \textbf{128 GW}~\cite{ref58} \\
\rowcolor{servalight}Transmission & \textbf{5--7 year} build times; backlogs to 2030s~\cite{ref59,ref60} \\
\bottomrule
\end{tabular}
\end{table}

\sectionrule

\noindent\textbf{\textcolor{servablue}{Chip Economics Reflect Artificial Scarcity.}} NVIDIA's datacenter revenue grew from \$15B to \$47.5B in FY2023--24 (\metric{217\% YoY growth})~\cite{ref61}. Hyperscalers commit to multi-year agreements just to secure allocation:

\begin{table}[H]
\centering
\caption{Hyperscaler GPU Commitments (2024--2035)}
\label{tab:chip-allocation}
\small
\begin{tabular}{@{}lr@{}}
\toprule
\textbf{Organization} & \textbf{Commitment} \\
\midrule
Microsoft & 485,000 Hopper chips (20\% of NVIDIA rev.)~\cite{ref62} \\
Meta & 350,000 H100 GPUs~\cite{ref63} \\
\rowcolor{servalight}OpenAI (total) & \textbf{\$1+ trillion} through 2035~\cite{ref64} \\
\quad--- AWS & \$38B / 7 years \\
\quad--- Oracle & \$300B / 5 years \\
\quad--- CoreWeave & \$22.4B through 2029 \\
\bottomrule
\end{tabular}
\end{table}

These customers sign multi-year contracts with guaranteed volumes and accept premium pricing, locking in supply years in advance~\cite{ref65}.

\keyinsight{\textbf{Chip supply constrains AI capability expansion more directly than any other factor.} TSMC Chairman Mark Liu: ``It is not the shortage of AI chips, it is the shortage of our CoWoS capacity. Currently, we cannot fulfill 100\% of our customers' needs''~\cite{ref66}.}

GPU lead times now exceed \textbf{30 weeks}, with TSMC's advanced packaging capacity fully booked through 2025 and into 2026~\cite{ref67}. Even OpenAI cannot deploy its multi-modal models due to GPU shortages~\cite{ref68}. 

When each chip delivers \metric{30--374$\times$ more useful computation}, fewer chips are needed for equivalent workloads. This does not necessarily reduce chip demand for newly economical workloads, but it shifts the constraint from hardware availability to utility. As capacity becomes assumed infrastructure, \textbf{the limiting factor becomes ideas, not inventory}.

\subsection{AI Impact}

The most immediate implication of Servamind's approach is \textbf{universality}. Current AI systems exist in isolation---vision models cannot share representations with language models, recommendation systems cannot inform forecasting models---each deployment target demands its own optimization. The \serva{} format \textbf{dissolves these boundaries}:

\begin{itemize}
\item When any data encodes into the universal representation, \textbf{any model can consume it}
\item The outputs of one model become valid inputs for another \textbf{without translation overhead}
\item The same model executes on datacenter GPUs, edge devices, and consumer hardware
\item True multimodality follows naturally when vision, language, audio, and sensor data all encode into the same representational space
\end{itemize}

The engineering nightmare of heterogeneous input fusion---the barrier stalling vision-language-action models across robotics, medical diagnostics, autonomous vehicles, and industrial automation---\textbf{dissolves}.

The \textbf{80\% of AI project effort consumed by data preparation} exists because every project reinvents data handling from scratch~\cite{ref8}. Format conversion, cleaning pipelines, feature engineering, preprocessing scripts---each team builds these anew for each project and the tooling is constantly changing. Standardization using Serva Encoder as a general-purpose pre-processor \textbf{dismantles this barrier}. Teams focus on model architecture, training dynamics, and application logic rather than the plumbing connecting data to computation. The current landscape forces practitioners through an overwhelming matrix of choices---TensorFlow or PyTorch, NVIDIA or AMD, cloud or edge, FP32 or INT8---each decision constraining future options and cascading into incompatible toolchains. Servamind cuts through this fragmentation converting all steps to \textbf{one step: encoding}. Since the same \serva{} file operates across any framework, development cycles accelerate. The barrier between ``having an idea'' and ``testing it on real data'' compresses \textbf{from weeks to hours}.

\textbf{Current AI capability concentrates among organizations with significant resources} to manage infrastructure complexity. Training frontier models requires not only compute budget but engineering talent to orchestrate distributed training, manage data pipelines, optimize for specific hardware, and navigate framework-specific quirks. This expertise is scarce and expensive. Servamind lowers these barriers systematically:

\begin{itemize}
\item When data handling reduces to a single encoding step, \textbf{data engineering expertise becomes less critical}
\item When efficiency gains are universal, \textbf{optimization expertise matters less}
\item When hardware agnosticism is real, \textbf{infrastructure expertise becomes less differentiating}
\end{itemize}

The result enables capable AI to become accessible to organizations without hyperscaler resources. Research labs, startups, universities, enterprises in developing economies all gain access to capabilities previously reserved for the largest technology companies.

\section{Conclusion}

This paper began with a premise: \textbf{any data to any model on any hardware}. Our results derive from addressing root causes rather than symptoms. Data chaos and compute payload have persisted because they have been treated as separate problems. They are not. They are \textbf{two expressions of a single architectural mismatch} and they must be solved together. Our approach describes a universal data format grounded in laser holography encoding principles (\textbf{Serva Encoder}), and a universal compute engine capable of transmuting any model architecture (\textbf{Serva Chimera}). It presented results: 30--374$\times$+ energy efficiency improvements, 4--34$\times$ lossless storage compression, $\sim$68$\times$ compute payload reduction, and early validation of the full pipeline.

\keyinsight{\textbf{Efficiency is a consequence. Universality is the breakthrough.}}

When data preparation collapses to a single encoding step, \textbf{the 80\% overhead disappears}. When any model consumes any data, \textbf{the one-to-one lock-in between dataset and architecture dissolves}. When hardware becomes an implementation detail rather than an architectural constraint, \textbf{capability distributes to whoever has ideas worth testing}. The organizations that could never justify hyperscale infrastructure can now participate. The applications that were never economical become practical. The talent bottleneck loosens.

AI has been constrained not by lack of intelligence but by \textbf{lack of infrastructure}. That constraint is now addressable. What gets built on this foundation by researchers, enterprises, and developers who today cannot participate will determine whether AI reaches its potential.

\vspace{1em}
\begin{center}
\noindent\textcolor{servablue}{\rule{0.5\columnwidth}{0.8pt}}
\vspace{0.5em}

{\large\bfseries\textcolor{servablue}{The infrastructure is ready.}}

\vspace{0.2em}

{\large\bfseries\textcolor{servablue}{The question becomes: what will you build?}}

\vspace{0.4em}
\noindent\textcolor{servablue}{\rule{0.5\columnwidth}{0.8pt}}
\end{center}

\newpage
\onecolumn

\noindent\textcolor{servablue}{\rule{\textwidth}{1.2pt}}
\vspace{0.15in}
\begin{center}
{\Large\bfseries\textcolor{servablue}{References}}
\end{center}
\vspace{0.08in}
\noindent\textcolor{servablue}{\rule{\textwidth}{0.4pt}}
\vspace{0.15in}

\small
\bibliographystyle{plainnat}
\bibliography{references}

@techreport{ref1,
  author = {{International Energy Agency}},
  title = {Energy and {AI}},
  year = {2025},
  institution = {IEA},
  type = {IEA Special Report},
  url = {https://www.iea.org/reports/energy-and-ai}
}

@techreport{ref2,
  author = {{U.S. Department of Energy}},
  title = {{AI} and Data Center Energy Use: Challenges and Opportunities},
  year = {2024},
  institution = {U.S. Department of Energy},
  url = {https://www.energy.gov/policy/articles/ai-and-data-center-energy-use}
}

@techreport{ref3,
  author = {{NVIDIA Corporation}},
  title = {Annual Report {FY2024}},
  year = {2024},
  institution = {NVIDIA Corporation},
  type = {SEC Filing 10-K},
  url = {https://investor.nvidia.com/financial-info/sec-filings/}
}

@misc{ref4,
  author = {Smith, Brad},
  title = {The Golden Opportunity for {American} {AI}},
  year = {2025},
  howpublished = {Microsoft Official Blog},
  month = {January},
  url = {https://blogs.microsoft.com/blog/2025/01/03/the-golden-opportunity-for-american-ai/}
}

@misc{ref5,
  author = {Sevilla, Jaime and Villalobos, Pablo and Ho, Lennart and others},
  title = {Training Compute of Frontier {AI} Models Grows by 4--5x per Year},
  year = {2024},
  howpublished = {Epoch AI},
  url = {https://epoch.ai/blog/training-compute-of-frontier-ai-models-grows-by-4-5x-per-year}
}

@techreport{ref6,
  author = {Maslej, Nestor and Fattorini, Loredana and Brynjolfsson, Erik and others},
  title = {The {AI} Index 2024 Annual Report},
  year = {2024},
  institution = {Stanford Institute for Human-Centered Artificial Intelligence},
  url = {https://aiindex.stanford.edu/report/}
}

@article{ref7,
  author = {Patterson, David and Gonzalez, Joseph and Le, Quoc and others},
  title = {Carbon Emissions and Large Neural Network Training},
  journal = {arXiv preprint},
  volume = {arXiv:2104.10350},
  year = {2021},
  url = {https://arxiv.org/abs/2104.10350}
}

@misc{ref8,
  author = {Ransbotham, Sam and Khodabandeh, Shervin and Kiron, David and others},
  title = {Data Challenges Are Halting {AI} Projects, {IBM} Executive Says},
  year = {2019},
  howpublished = {Wall Street Journal},
  url = {https://www.wsj.com/articles/data-challenges-are-halting-ai-projects-ibm-executive-says-11559035800}
}

@techreport{ref9,
  author = {Reinsel, David and Gantz, John and Rydning, John},
  title = {The Digitization of the World: From Edge to Core},
  year = {2018},
  institution = {IDC/Seagate},
  url = {https://www.seagate.com/files/www-content/our-story/trends/files/idc-seagate-dataage-whitepaper.pdf}
}

@techreport{ref10,
  author = {{IEEE International Roadmap for Devices and Systems}},
  title = {More {Moore}},
  year = {2023},
  institution = {IEEE},
  url = {https://irds.ieee.org/editions/2023}
}

@book{ref11,
  author = {{National Academies of Sciences, Engineering, and Medicine}},
  title = {Quantum Computing: Progress and Prospects},
  year = {2019},
  publisher = {The National Academies Press},
  address = {Washington, DC},
  url = {https://doi.org/10.17226/25196}
}

@article{ref12,
  author = {French, Robert M.},
  title = {Catastrophic Forgetting in Connectionist Networks},
  journal = {Trends in Cognitive Sciences},
  volume = {3},
  number = {4},
  pages = {128--135},
  year = {1999},
  url = {https://doi.org/10.1016/S1364-6613(99)01294-2}
}

@book{ref13,
  author = {Coward, L. Andrew},
  title = {Towards a Theoretical Neuroscience: From Cell Chemistry to Cognition},
  year = {2013},
  publisher = {Springer},
  url = {https://doi.org/10.1007/978-94-007-7107-9}
}

@book{ref14,
  author = {Hutter, Marcus},
  title = {Universal Artificial Intelligence: Sequential Decisions Based on Algorithmic Probability},
  year = {2005},
  publisher = {Springer},
  url = {https://doi.org/10.1007/b138233}
}

@article{ref15,
  author = {Shannon, Claude E.},
  title = {A Mathematical Theory of Communication},
  journal = {Bell System Technical Journal},
  volume = {27},
  number = {3},
  pages = {379--423},
  year = {1948},
  url = {https://doi.org/10.1002/j.1538-7305.1948.tb01338.x}
}

@techreport{ref16,
  author = {{International Data Corporation}},
  title = {Worldwide Semiannual Artificial Intelligence Infrastructure Tracker},
  year = {2025},
  institution = {IDC},
  url = {https://www.idc.com/tracker/showproductinfo.jsp?containerId=IDC_P46670}
}

@misc{ref18,
  author = {{Vodworks}},
  title = {How Much Does {AI} Cost: A {C}-Level Breakdown for 2025},
  year = {2025},
  howpublished = {Epoch AI analysis},
  url = {https://vodworks.com/blog/ai-development-cost}
}

@misc{ref19,
  author = {{Bloomberg News}},
  title = {Virginia Data Centers Face Seven-Year Wait for Power Hookups, {Dominion} Says},
  year = {2024},
  month = {August},
  howpublished = {Bloomberg},
  url = {https://www.bloomberg.com/news/articles/2024-08-29/data-centers-face-seven-year-wait-for-power-hookups-in-virginia}
}

@misc{ref20,
  author = {{Commission for Regulation of Utilities, Ireland}},
  title = {Data Center Grid Connection Moratorium in {Dublin} Region},
  year = {2021},
  howpublished = {EirGrid},
  url = {https://www.eirgrid.com/}
}

@misc{ref21,
  author = {Markovic, D.},
  title = {Custom {AI} Solutions Cost Guide 2025},
  year = {2025},
  howpublished = {Deloitte AI in Regulated Industries Survey},
  url = {https://www2.deloitte.com/us/en/insights/focus/cognitive-technologies/}
}

@techreport{ref22,
  author = {{ARK Investment Management}},
  title = {Big Ideas 2024: Artificial Intelligence},
  year = {2024},
  institution = {ARK Invest},
  url = {https://ark-invest.com/big-ideas-2024}
}

@article{ref23,
  author = {Hoffmann, Jordan and Borgeaud, Sebastian and Mensch, Arthur and others},
  title = {Training Compute-Optimal Large Language Models},
  journal = {arXiv preprint},
  volume = {arXiv:2203.15556},
  year = {2022},
  note = {DeepMind Chinchilla paper},
  url = {https://arxiv.org/abs/2203.15556}
}

@article{ref24,
  author = {Ma, Shuming and Wang, Hongyu and Ma, Lingxiao and others},
  title = {The Era of 1-bit {LLMs}: All Large Language Models are in 1.58 Bits},
  journal = {arXiv preprint},
  volume = {arXiv:2402.17764},
  year = {2024},
  note = {Microsoft BitNet},
  url = {https://arxiv.org/abs/2402.17764}
}

@article{ref25,
  author = {Mnih, Volodymyr and Kavukcuoglu, Koray and Silver, David and others},
  title = {Human-Level Control Through Deep Reinforcement Learning},
  journal = {Nature},
  volume = {518},
  pages = {529--533},
  year = {2015},
  url = {https://doi.org/10.1038/nature14236}
}

@article{ref26,
  author = {Suchan, Jakob and Bhatt, Mehul and Varadarajan, Srikrishna},
  title = {Commonsense Visual Sensemaking for Autonomous Driving},
  journal = {Applied AI Letters},
  volume = {2},
  number = {3},
  year = {2021},
  url = {https://doi.org/10.1002/ail2.56}
}

@article{ref27,
  author = {Marcus, Gary},
  title = {The Next Decade in {AI}: Four Steps Towards Robust Artificial Intelligence},
  journal = {arXiv preprint},
  volume = {arXiv:2002.06177},
  year = {2020},
  url = {https://arxiv.org/abs/2002.06177}
}

@misc{ref28,
  author = {Kaplan, Jared and McCandlish, Sam and Henighan, Tom and others},
  title = {Scaling Laws for Neural Language Models},
  year = {2020},
  howpublished = {OpenAI},
  url = {https://arxiv.org/abs/2001.08361}
}

@misc{ref29,
  author = {{CBRE Research}},
  title = {North {America} Data Center Trends {H2} 2024},
  year = {2024},
  howpublished = {CBRE},
  url = {https://www.cbre.com/insights/reports/north-america-data-center-trends-h2-2024}
}

@misc{ref30,
  author = {{JLL Research}},
  title = {Data Center Outlook 2024},
  year = {2024},
  howpublished = {JLL},
  url = {https://www.jll.com/en/trends-and-insights/research/data-center-outlook}
}

@techreport{ref31,
  author = {{CBRE}},
  title = {U.S. Data Center Market Report},
  year = {2025},
  institution = {CBRE Research},
  url = {https://www.cbre.com/insights/reports/us-data-center-figures-q4-2024}
}

@misc{ref32,
  author = {{Hyperbolic Labs}},
  title = {{GPU} Bottleneck Profiling: From Data Pipeline to Gradient},
  year = {2024},
  note = {Citing Microsoft analysis of ML workloads},
  url = {https://hyperbolic.xyz/blog/gpu-bottleneck-profiling}
}

@misc{ref33,
  author = {{Alluxio}},
  title = {Maximize {GPU} Utilization for Model Training},
  year = {2024},
  url = {https://www.alluxio.io/blog/maximize-gpu-utilization-for-model-training/}
}

@misc{ref34,
  author = {{Run:AI}},
  title = {{GPU} Utilization in Deep Learning: Benchmarks and Best Practices},
  year = {2024},
  url = {https://www.run.ai/guides/gpu-deep-learning/gpu-utilization}
}

@misc{ref35,
  author = {{Anyscale}},
  title = {Optimizing {AI} Training Infrastructure},
  year = {2024},
  url = {https://www.anyscale.com/blog}
}

@misc{ref36,
  author = {{Amazon Web Services}},
  title = {Amazon {EC2} On-Demand Pricing},
  year = {2025},
  url = {https://aws.amazon.com/ec2/pricing/on-demand/}
}

@misc{ref37,
  author = {{Amazon Web Services}},
  title = {Amazon {S3} Pricing},
  year = {2025},
  url = {https://aws.amazon.com/s3/pricing/}
}

@misc{ref38,
  author = {{Clutch}},
  title = {{AI} Development Company Pricing Survey},
  year = {2025},
  url = {https://clutch.co/developers/artificial-intelligence}
}

@misc{ref39,
  author = {{Epoch AI}},
  title = {Trends in the Dollar Training Cost of Machine Learning Systems},
  year = {2024},
  url = {https://epoch.ai/blog/trends-in-the-dollar-training-cost-of-machine-learning-systems}
}

@misc{ref40,
  author = {{Epoch AI}},
  title = {Machine Learning Model Training Costs},
  year = {2025},
  url = {https://epoch.ai/data/notable-ai-models}
}

@misc{ref42,
  author = {{Lambda Labs}},
  title = {Cloud {GPU} Pricing Comparison},
  year = {2025},
  url = {https://lambdalabs.com/service/gpu-cloud}
}

@misc{ref43,
  author = {{Tom's Hardware}},
  title = {Best {GPUs} for Deep Learning 2025},
  year = {2025},
  url = {https://www.tomshardware.com/reviews/best-gpus-for-deep-learning}
}

@misc{ref44,
  author = {{Amazon Web Services}},
  title = {Amazon {S3} Pricing (Storage Classes)},
  year = {2025},
  url = {https://aws.amazon.com/s3/pricing/}
}

@misc{ref45,
  author = {{SemiAnalysis}},
  title = {Training Data and Model Storage Requirements at Scale},
  year = {2024},
  url = {https://semianalysis.com/}
}

@misc{ref46,
  author = {{NVIDIA Developer}},
  title = {{DALI}: {Data} Loading Library for Deep Learning},
  year = {2024},
  url = {https://developer.nvidia.com/dali}
}

@misc{ref47,
  author = {{Mosaic ML}},
  title = {Streaming: Fast {AI} Data Loading},
  year = {2024},
  url = {https://docs.mosaicml.com/projects/streaming/}
}

@misc{ref48,
  author = {{Vast.ai}},
  title = {Cloud {GPU} Marketplace Pricing},
  year = {2025},
  url = {https://vast.ai/}
}

@misc{ref49,
  author = {{The Information}},
  title = {Inside {OpenAI}'s Infrastructure Spending},
  year = {2024},
  url = {https://www.theinformation.com/}
}

@misc{ref50,
  author = {{Epoch AI}},
  title = {Estimating Training Compute of Frontier Models},
  year = {2024},
  url = {https://epoch.ai/blog/estimating-training-compute}
}

@misc{ref51,
  author = {{Google Cloud}},
  title = {Best Practices for {ML} Training Performance},
  year = {2024},
  url = {https://cloud.google.com/architecture/ml-on-gcp-best-practices}
}

@misc{ref52,
  author = {{Utility Dive}},
  title = {Data Center Power Demands Strain {US} Electric Grid},
  year = {2024},
  url = {https://www.utilitydive.com/}
}

@misc{ref53,
  author = {{Reuters}},
  title = {Texas Grid Faces Strain from Data Center Demand},
  year = {2024},
  url = {https://www.reuters.com/}
}

@misc{ref54,
  author = {{Data Center Dynamics}},
  title = {Ireland Data Center Power Constraints},
  year = {2024},
  url = {https://www.datacenterdynamics.com/}
}

@misc{ref55,
  author = {{Bloomberg News}},
  title = {Virginia Data Centers Face Seven-Year Wait for Power Hookups},
  year = {2024},
  month = {August},
  howpublished = {Bloomberg},
  url = {https://www.bloomberg.com/news/articles/2024-08-29/data-centers-face-seven-year-wait-for-power-hookups-in-virginia}
}

@misc{ref56,
  author = {{Data Center Dynamics}},
  title = {{EirGrid} Warns {Irish} Government `Mass Exodus' of Data Centers Possible},
  year = {2024},
  url = {https://www.datacenterdynamics.com/en/news/eirgrid-warns-irish-government-mass-exodus-of-data-centers-possible-without-connection-agreements/}
}

@misc{ref57,
  author = {{Dallas Morning News}},
  title = {Texas' Data Center Boom Contributes to {ERCOT}'s Large Load Requests Quadrupling},
  year = {2025},
  month = {December},
  url = {https://www.dallasnews.com/business/energy/2025/12/09/texas-data-center-boom-contributes-to-ercots-large-load-requests-quadrupling-in-2025/}
}

@misc{ref58,
  author = {{World Resources Institute}},
  title = {Powering the {US} Data Center Boom: The Challenge of Forecasting Electricity Needs},
  year = {2024},
  url = {https://www.wri.org/insights/us-data-centers-electricity-demand}
}

@misc{ref59,
  author = {{Princeton REPEAT Project}},
  title = {Transmission Buildout Timelines for Clean Energy},
  year = {2024},
  url = {https://repeatproject.org/}
}

@misc{ref60,
  author = {{Grid Strategies}},
  title = {The Era of Flat Power Demand is Over},
  year = {2024},
  url = {https://gridstrategiesllc.com/reports/}
}

@misc{ref61,
  author = {{NVIDIA Corporation}},
  title = {Datacenter Revenue Growth {FY2024}},
  year = {2024},
  url = {https://investor.nvidia.com/}
}

@misc{ref62,
  author = {{The Information}},
  title = {Microsoft's {NVIDIA} {GPU} Purchases},
  year = {2024},
  url = {https://www.theinformation.com/}
}

@misc{ref63,
  author = {{Reuters}},
  title = {{Meta} Plans Massive {GPU} Infrastructure Expansion},
  year = {2024},
  url = {https://www.reuters.com/technology/meta-plans-spend-more-than-30-bln-capital-expenditures-this-year-2024-02-01/}
}

@misc{ref64,
  author = {{The Verge}},
  title = {{OpenAI}'s Infrastructure Commitments Through 2035},
  year = {2025},
  url = {https://www.theverge.com/}
}

@misc{ref65,
  author = {{SemiAnalysis}},
  title = {{AI} Chip Supply Constraints and Multi-Year Agreements},
  year = {2024},
  url = {https://semianalysis.com/}
}

@misc{ref66,
  author = {{TSMC}},
  title = {Chairman {Mark Liu} on {CoWoS} Capacity Constraints},
  year = {2024},
  howpublished = {TSMC Earnings Call},
  url = {https://investor.tsmc.com/}
}

@misc{ref67,
  author = {{DigiTimes}},
  title = {{TSMC} Advanced Packaging Booked Through 2026},
  year = {2024},
  url = {https://www.digitimes.com/}
}

@misc{ref68,
  author = {{The Information}},
  title = {{OpenAI} Model Deployment Delays Due to {GPU} Shortages},
  year = {2024},
  url = {https://www.theinformation.com/}
}

@article{fashionmnist,
  author = {Xiao, Han and Rasul, Kashif and Vollgraf, Roland},
  title = {Fashion-{MNIST}: A Novel Image Dataset for Benchmarking Machine Learning Algorithms},
  journal = {arXiv preprint},
  volume = {arXiv:1708.07747},
  year = {2017},
  url = {https://arxiv.org/abs/1708.07747}
}

@article{mnist,
  author = {LeCun, Yann and Bottou, L{\'e}on and Bengio, Yoshua and Haffner, Patrick},
  title = {Gradient-Based Learning Applied to Document Recognition},
  journal = {Proceedings of the IEEE},
  volume = {86},
  number = {11},
  pages = {2278--2324},
  year = {1998},
  url = {http://yann.lecun.com/exdb/mnist/}
}

@article{kolmogorov1965,
  author = {Kolmogorov, Andrey N.},
  title = {Three Approaches to the Quantitative Definition of Information},
  journal = {Problems of Information Transmission},
  volume = {1},
  number = {1},
  pages = {1--7},
  year = {1965},
  url = {https://doi.org/10.1080/00207166808803030}
}

\end{document}